\documentclass[journal,onecolumn,draftclsnofoot,11pt]{IEEEtran}

\usepackage[noadjust]{cite}

\ifCLASSINFOpdf
   \usepackage[pdftex]{graphicx}
   \graphicspath{{../pdf/}{../jpeg/}}
   \DeclareGraphicsExtensions{.pdf,.jpeg,.png}
\else
   \usepackage[dvips]{graphicx}
   \graphicspath{{../eps/}}
   \DeclareGraphicsExtensions{.eps}
\fi

\usepackage{psfrag}
\usepackage{tikz}
\usepackage{pgfplots}
\usetikzlibrary{shapes.misc}
\usetikzlibrary{arrows}
\usepackage[labelformat=simple]{subcaption}

\usepackage{amsmath}
\usepackage{amssymb}
\usepackage{bm}

\usepackage[ruled,linesnumbered]{algorithm2e}

\usepackage{amsthm}

\usepackage{xfrac}
\usepackage[font=footnotesize,labelsep=period]{caption}

\usepackage{tabularx, booktabs}
\usepackage{longtable}
\usepackage{enumitem}
\usepackage{epstopdf}
\usepackage{xcolor, colortbl}
\usepackage{multirow}

\usepackage{blindtext}
\usepackage[utf8]{inputenc}

\newcommand\numberthis{\addtocounter{equation}{1}\tag{\theequation}}

\begin{document}

\title{On the Beamformed Broadcast Signaling for Millimeter Wave Cell Discovery: Performance Analysis and Design Insight}

\author{Yilin~Li,
        Jian~Luo,
        Mario H. Casta$\tilde{\text{n}}$eda Garcia,
        Richard A. Stirling-Gallacher,
        Wen~Xu,
        and~Giuseppe~Caire
\thanks{Yilin Li, Jian Luo, Mario H. Casta$\tilde{\text{n}}$eda Garcia, Richard A. Stirling-Gallacher, and Wen Xu are with Huawei Technologies D{\"u}sseldorf GmbH, German Research Center, 80992 Munich, Germany (emails: yilin.li@huawei.com, jianluo@huawei.com, mario.castaneda@huawei.com, richard.sg@huawei.com, wen.dr.xu@huawei.com). Giuseppe Caire is with the Communications and Information Theory Group, Technische Universit{\"a}t Berlin, 10587 Berlin, Germany, and with the Department of Electrical Engineering, University of Southern California, Los Angeles, CA 90089, USA (email: caire@tu-berlin.de)}
}

\maketitle

\begin{abstract}
Availability of abundant spectrum has enabled millimeter wave (mm-wave) as a prominent candidate solution for the next generation cellular networks. Highly directional transmissions are essential for exploitation of mm-wave bands to compensate high propagation loss and attenuation. The directional transmission, nevertheless, necessitates a specific design for mm-wave initial cell discovery, as conventional omni-directional broadcast signaling may fail in delivering the cell discovery information. To address this issue, this paper provides an analytical framework for mm-wave beamformed cell discovery based on an information theoretical approach. Design options are compared considering four fundamental and representative broadcast signaling schemes to evaluate discovery latency and signaling overhead. The schemes are then simulated under realistic system parameters. Analytical and simulation results reveals four key findings: (i) For cell discovery without knowledge of beacon timing, analog/hybrid beamforming performs as well as digital beamforming in terms of cell discovery latency; (ii) Single beam exhaustive scan optimize the latency, however leads to overhead penalty; (iii) Multi-beam simultaneous scan can significantly reduce the overhead, and provide the flexibility to achieve trade-off between the latency and the overhead; (iv) The latency and the overhead are relatively insensitive to extreme low block error rates.
\end{abstract}

\section{Introduction}
%
%
%
%
\IEEEPARstart{M}{illimeter} wave (mm-wave) frequency bands between 6 and 100 GHz have drawn significant attention for the next generation cellular communication systems~\cite{Rangan}\cite{Rappaport}, where the available bandwidths are much wider than today{\rq}s cellular allocations~\cite{Pi}\cite{YLiCommag}. Mm-wave signals, however, suffer from increased isotropic free space loss, higher penetration loss, and propagation attenuation, resulting in outages and intermittent channel quality~\cite{Alejos}. In this regard, enhanced antenna gain is required at both transceiver sides to completely compensate the loss and the attenuation of mm-wave transmissions. 

Fortunately, the very small wavelengths of the mm-wave signals, combined with advanced low power  CMOS RF circuits, enable the deployment of large-scale miniaturized antennas and the exploitation of beamforming and spatial multiplexing~\cite{Akdeniz}. As a result, reliance of highly directional transmission and reception considerably complicates initial cell discovery in mm-wave cellular communications. While conventional cellular systems, such as 3GPP LTE/LTE-A~\cite{3GPPOD, 3GPPMACPS, 3GPPRRCPS}, support multi-antenna diversity techniques and spatial multiplexing with beamforming, underlying design assumption is that the initial cell discovery can be conducted entirely with omni-directional transmissions or transmissions in fixed antenna patterns~\cite{Barati}. LTE base station (BS), for example, generally does not apply beamforming when transmitting synchronization and broadcasting signals. Directional transmissions are typically exploited only after initial access has been established.

\begin{figure}[htb]
	\centering
	\includegraphics [scale=0.8]{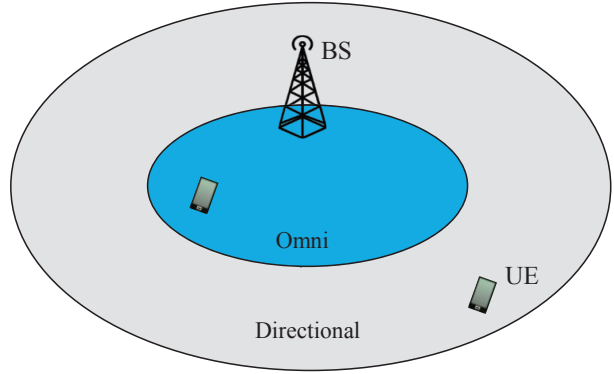}
	\caption{Illustration of mismatch between discoverable area and achievable area of mm-wave communication networks. Due to highly directional transmissions, mm-wave broadcast signaling exhibits different cell range in omni-directional transmission (blue) vs. directional transmission (gray). User equipment (UE) locates at gray area is not able to establish communication link with the BS.}\label{Figure1}
\end{figure}

Moreover, for mm-wave communications, omni-directional broadcast signaling transmission may fail in the initial cell discovery procedure, as utilization of high directional antenna would create mismatch between discoverable range and achievable range~\cite{QCLi}. Specifically, for systems operating at mm-wave bands, applying conventional cell discovery technique would lead to a smaller discoverable area than the achievable area (the area where reasonable data rates can be achieved), as illustrated in Fig.~\ref{Figure1}. Therefore, the initial cell discovery procedure is expected to be designed properly to establish communication links via directional transmission and exploit resources in spatial dimension.

\subsection{Related Works}

Cell discovery, particularly at mm-wave bands, has been thoroughly investigated by recent research efforts~\cite{Shokri-GhadikolaeiICC15, BLi, Jeong, Shokri-GhadikolaeiTC15, Palacios, DeDonno, Giordani, Barati, YingzheLi, YLiVTC} and standards associations~\cite{802.11ad, Nitsche2014, 802.15.3c}. Deafness, as an explicit consequence of directional transmission and reception, emerges when main beam of transmitter (Tx) and intended beam of receiver (Rx) are not aligned~\cite{Shokri-GhadikolaeiICC15}. To address this issue, communication links are established by beam scan procedure~\cite{BLi, Jeong}, in which an exhaustive scan over all possible combinations of transmission and reception directions is performed through a sequence of broadcasting signals. Leveraging synchronization from macro BSs, authors in~\cite{Shokri-GhadikolaeiTC15} show that the cell discovery efficiency can be enhanced with sequential spatial search.

Incorporating beam scan concept facilitates beamforming procedure, however introduces alignment latency that is the time of matching beam pairs to complete cell discovery. Current standardization activities, e.\,g. wireless local area network (WLAN) standard 802.11ad~\cite{802.11ad} and wireless personal area network (WPAN) standard 802.15.3c~\cite{802.15.3c}, have proposed exhaustive search of transceiver beams and a two-stage hierarchical beam scan technique to reduce the alignment latency. Specifically, a coarse sector-level beam alignment is performed, followed by a beam-level refinement phase. However, such schemes suffer from high latency required for matching the beam pairs. Authors of~\cite{Palacios} and~\cite{DeDonno} proposed an enhanced beam codebook design for better alignment of the beams.

The exhaustive and hierarchical strategies are compared in~\cite{Giordani}, which shows that hierarchical search generally has smaller cell discovery access delay, but exhaustive search gives better coverage to cell-edge users. Further enhancements on the initial beam
training design are proposed in~\cite{Barati}, which has observed that the cell discovery delay can be reduced by omni-directional transmission from the BSs during cell search. Authors in~\cite{YingzheLi} proposed an initial access framework including several beam paring protocols to investigate the trade-off between initial access delay and user-perceived downlink throughput.

Furthermore, geometry information has been recently recognized as a promising candidate to improve
the mm-wave cell discovery efficiency~\cite{Filippini, Capone, Bai, Alkhateeb}. By incorporating context information related to user position, authors in~\cite{Filippini} and~\cite{Capone} show that the performance gain of cell discovery can be improved. Similarly, authors in~\cite{Bai} and~\cite{Alkhateeb} demonstrated that mm-wave networks with exploitation of statistical blockage models can achieve comparable coverage and considerable data rate than conventional cellular networks.

The aforementioned research studies are quite inspiring. However, they are either a point-to-point communication (\cite{Giordani, Filippini, Capone}), or only consider one typical user with a few nearby BSs (\cite{Bai, Alkhateeb, Shokri-GhadikolaeiTC15}). Some assume that the association between user and its serving BS has already been established (\cite{Bai, Alkhateeb}), while in fact the initial access is a key challenge in mm-wave cellular communications. Moreover, system-level analysis of mm-wave cell discovery has not to be considered\cite{BLi, Barati}).

\subsection{Contributions}
In this paper, we consider the fact that in initial access phase, rough beam alignment would be sufficient, and aligned beam refinement can be done in later stages via scheduled resources. The most important task of broadcast signaling during initial access is to convey necessary information for cell discovery to UEs in the most efficient way. Therefore, we investigate four broadcast signaling schemes including an exhaustive beam scan scheme as the baseline, and evaluate the performance of mm-wave cell discovery incorporating the impact of broadcast signaling, as well as draw some insights into the design of mm-wave cell discovery. The contributions of this paper are summarized as follows.

\begin{itemize}
	\item Development of an accurate analytical framework for mm-wave cell discovery under various broadcast signaling schemes. The framework divides the system-operation into several cases depending on knowledge of beacon timing and beacon interval integration. These cases are shown to result in different \textit{cell discovery latencies} (CDLs) and \textit{signaling overheads} (OHs). In particular, for the case without knowledge of beacon timing, it is demonstrated that analog/hybrid beamforming performs as well as digital beamforming in terms of the CDL.
	\item Comparison of the CDL and the OH. The baseline scheme, i.\,e. single beam exhaustive scan, leads to the best CDL performance, which is the opposite to an intuitive consideration that multi-beam simultaneous scan can accelerate the discovery. This scheme, however, causes high OH due to the impact of guard interval (GI) reserved for beam switching. By contrast, other schemes that apply multi-beam simultaneous scan generally provides the flexibility to achieve a trade-off between latency and OH by configuring the number of simultaneous beams.
	\item A detailed system-level performance evaluation for mm-wave cell discovery. Different from the link-level analysis in~\cite{Giordani, Barati, Filippini, Alkhateeb}, we derive system-level performance metrics to capture the cell discovery efficiency, including the CDL and the OH. Our analytical results are validated against the detailed system-level simulations.
	\item Leveraging the analysis and the simulation results to provide insights into the answers of the key questions: (i) How wide should the beam be? (ii) Is it beneficial to exploit multi-beam simultaneous scan? (iii) If so, how many simultaneous beams should be exploited? (iv) What is the impact of block error rate?
\end{itemize}

The remainder of the paper is organized as follows: Section II presents the system model and Section III compares and analyzes the different broadcast signaling schemes. Section IV presents the analysis on the cell discovery procedure. The analysis is then verified by extensive simulations in Section V, followed by a summary concluding the paper in Section VI.

A conference version of this paper has appeared in~\cite{YLiVTC}. This paper includes all derivations, discussion of extensions, more detailed and extensive simulations, and has taken into account more options for beacon interval integration.

\section{System Model}
In this section, we introduce the network, propagation, antenna, and beam scan models considered for evaluating the mm-wave cell discovery performance in this paper. The important notations and system parameters defined in this section are summarized in Table~\ref{Table1} and will be used in the rest of this paper.

\begin{table}[htb]
	\centering
	\caption{System Model Parameters}
	\label{Table1}
	\resizebox{340pt}{!}{%
	\begin{tabular}{|l|p{80mm}|p{35mm}|}
		\hline \\[-1em]
		Notation & Description & Value \\ \hline \\[-1em]
		$r$ & Cell radius & $100$ m \\ \hline \\[-1em]
		$h_\text{AP}$ & Height of AP antennas & $15$ m \\ \hline \\[-1em]
		$h_\text{UE}$ & Height of UE access plane & $1.5$ m \\ \hline \\[-1em]
		$p{(d)}$ & Probability of a link with distance $d$ to be LOS &  \\ \hline \\[-1em]
		$d_1$ & Parameters in $d_{1}/d_{2}$ model & $20$ \\ \hline \\[-1em]
		$d_2$ & Parameters in $d_{1}/d_{2}$ model & $39$ \\ \hline \\[-1em]
		$L(f,d)$ & Pathloss of a link at carrier frequency $f$ and distance $d$ &  \\ \hline \\[-1em]
		$f$ & Carrier frequency & $28\, \text{GHz}$ \\ \hline \\[-1em]
		$c$ & Speed of light & $3\times10^8$ m/s \\ \hline \\[-1em]
		$n_\text{L}$ & Pathloss exponent & LOS, NLOS: $2.1,3.17$ \\ \hline \\[-1em]
		SF & Shadowing factor & LOS, NLOS: $3.76,8.09$ \\ \hline \\[-1em]
		$N_\text{Tx}$ & Number of antennas at AP & $128$ \\ \hline \\[-1em]
		$M_\text{Tx}$ & Number of RF chains at AP & $128$ \\ \hline \\[-1em]
		$N$ & Number of beam scan areas & $1,2,4,...,128$ \\ \hline \\[-1em]
		$M$ & Number of simultaneous beams & Divisor of $N$ \\ \hline \\[-1em]
		$N_\text{H}$ & Number of beam scan areas in horizontal dimension & $1,2,4,8,16$ \\ \hline \\[-1em]
		$N_\text{V}$ & Number of beam scan areas in vertical dimension & $1,2,4,8$ \\ \hline \\[-1em]
		$M_\text{H}$ & Number of simultaneous beams in horizontal dimension & Divisor of $N_\text{H}$ \\ \hline \\[-1em]
		$M_\text{V}$ & Number of simultaneous beams in vertical dimension & Divisor of $N_\text{V}$ \\ \hline \\[-1em]
		$S$ & Number of beam slots & $N/M$ \\ \hline \\[-1em]
		$G$ & Antenna beamforming gain & \\ \hline \\[-1em]
		$d_\text{H}$ & Height of approximated rectangular beam pattern & $d_\text{H}=d\sin{\theta}$ \\ \hline \\[-1em]
		$d_\text{V}$ & Width of approximated rectangular beam pattern & $d_\text{V}=d\sin{\phi}$ \\ \hline \\[-1em]
		$\theta$ & Elevation angle of Tx beam & $\sfrac{2\pi}{N_\text{H}}$ \\ \hline \\[-1em]
		$\phi$ & Azimuth angle of Rx beam & $\arctan(\frac{r}{h_\text{AP}-h_\text{UE}})/{N_\text{V}}$ \\ \hline \\[-1em]
		$T$ & Frame length & $200$ $\mu$s \\ \hline
	\end{tabular}}
\end{table}

\subsection{Network Model}
\begin{figure}[htb]
	\centering
	\includegraphics [scale=0.8]{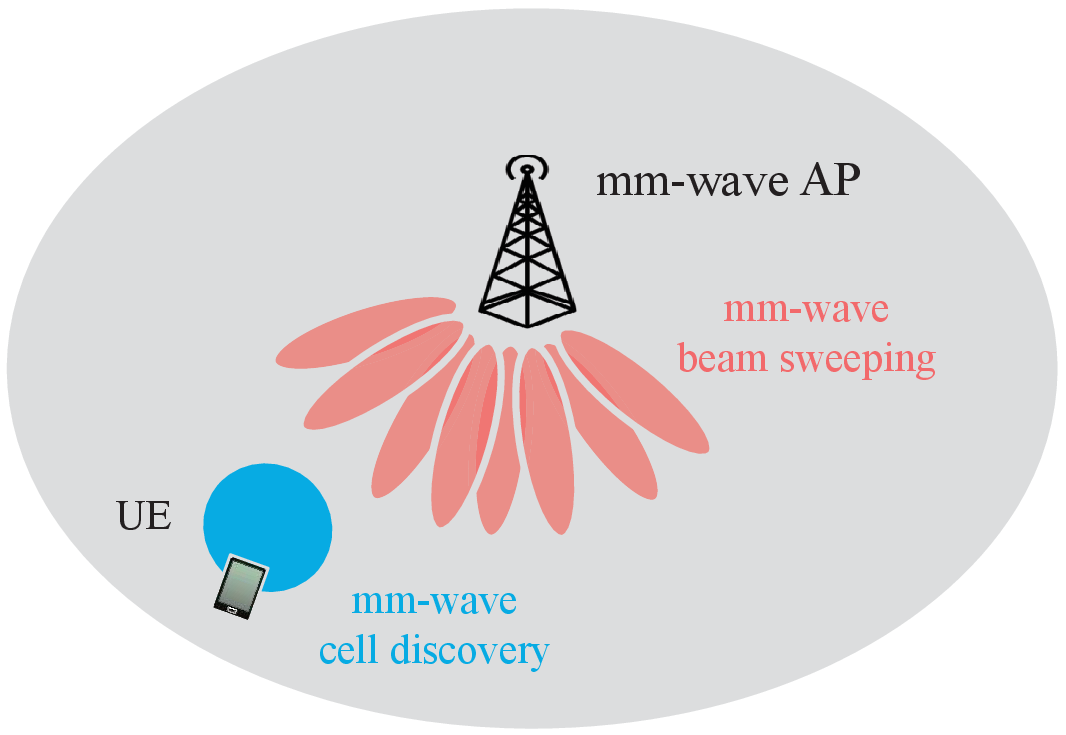}
	\caption{Illustration of a single cell mm-wave cellular network with one AP. Only one UE is shown.}\label{Figure2}
\end{figure}

Without loss of generality, we envision a single cell downlink cellular network with one mm-wave access point (AP) at its center of radius $r=100$ m. In general, the AP can be located indoor or outdoor. In this paper, we focus on the performance of mm-wave cellular networks with outdoor BS. UEs are also assumed to be outdoor and randomly ``dropped'' in the network. Fig.~\ref{Figure2} shows an example of the considered mm-wave cellular network.

\subsection{Blockage and Propagation Model}
We assume that buildings act as propagation blockages for the considered outdoor communication between the AP and the UEs. Based on that, the link state between the AP and each UE is determined to be either line-of-sight (LOS) or non-line-of-sight (NLOS) by considering whether any buildings/walls intersect the direct path between Tx (AP) and Rx (UE). To incorporate the LOS/NLOS state into our system model, we adopt the $d_1$/$d_2$ model in~\cite{3GPPSCM, ITUR}:
\begin{equation} 
p(d) = \min(\frac{d_1}{d},1)(1-e^{-{d}/{d_2}}) + e^{-{d}/{d_2}}, \label{eqnLOSNLOS}
\end{equation}
where the link between the Tx and the Rx at distance $d$ is determined to be LOS or NLOS according to the LOS probability $p(d)$. 

The path loss for a link with distance $d$ in dB operating at carrier frequency $f$ is given by~\cite{RappaportTC}:
\begin{equation} 
L(f,d) = 20\log_{10}{\frac{4{\pi}f}{c}} + 10n_\text{L}\log_{10}{d} + \text{SF}, \label{eqnPL}
\end{equation}
where $f$ is the carrier frequency in Hz, $n$ is the path loss exponent, and $d$ is the distance between the Tx and the Rx in meters. The first term of~\eqref{eqnPL} indicates the free space path loss at $1$ m, where $c$ is the speed of light. Pathloss exponent is represented by $n_\text{L}$. Impact of the objects such as trees, cars, furniture, etc. is not represented in the blockage model and is modeled separately using the shadowing factor SF in dB. Further, an additive white Gaussian noise (AWGN) channel is assumed within each beam.

It is worth noting that throughout this work we assume a standalone mm-wave network. Some recent papers such as~\cite{Nitsche2015}~\cite{Marcano} discuss legacy bands (sub-6 GHz) assisted cell discovery, proposing joint search for UE between mm-wave small cell and macro cell. However, we still provide the performance analysis of cell discovery in a non-standalone network (see Section IV-B), where the existence of legacy band refers to the support of synchronization between AP and UE, rather than the joint search between networks.

In addition, authors in~\cite{Capone, Jung, Abbas} take advantage of “external localization service” to get positioning information provided by e.\,g. using GPS. While the cell discovery in most of these works requires LOS paths, we also take into account establishing NLOS links. Therefore, in our simulation, we evaluate the cell discovery performance with a realistic LOS/NLOS channel model.

\subsection{Antenna and Beamforming Model}
\begin{figure}[htb]
	\centering
	\includegraphics[scale=0.8]{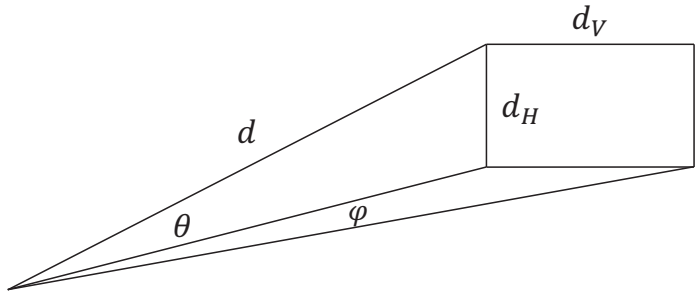}
	\caption{Tx beamforming gain is calculated by a rectangular sectorized pattern.}
	\label{Figure3}
\end{figure}

We assume that the AP employs directional steerable antenna arrays of $N_\text{Tx}$ antennas and can perform both 2D and 3D beamforming. The AP is equipped with $M_\text{Tx}$ RF chains, such that multiple simultaneous beams can be transmitted. In case of 3D beamforming, Tx beams are scanning over both horizontal ($[0,2\pi]$) and vertical ($[0,\arctan(\frac{r}{h_\text{AP}-h_\text{UE}})]$) directions, where $r$, $h_\text{AP}$, and $h_\text{UE}$ are the cell radius, the height of AP antennas, and the height of access plane of UE, respectively. The UEs are assumed to be able to synthesize quasi-omni antenna patter (e.\,g. as in 802.11ad~\cite{802.11ad}) for signal reception. An extension to beam-scanning reception is possible and left as future work. For analytical tractability, we assume the actual antenna pattern is approximated by a sectorized beam pattern (\cite{Bai, Alkhateeb, Shokri-GhadikolaeiTC15}) depicted in Fig. 1(b) of~\cite{Alkhateeb}, where we calculate the Tx array beamforming gain by a rectangular sectorized pattern as
\begin{equation} 
G = \frac{\text{Area of cell sphere}}{\text{Area of rectangular}} = \frac{4{\pi}d^2}{d_\text{H}d_\text{V}} = \frac{4{\pi}}{\sin{\theta}\sin{\phi}},
\label{eqnAntennaGain}
\end{equation}
where $\theta$ and $\phi$ represent the elevation and the azimuth angle of a beam. The approximated sectorized area (the rectangular) of height $d_\text{H}$ and width $d_\text{V}$ can be obtained as $d_\text{H}=d\sin{\theta}$ and $d_\text{V}=d\sin{\phi}$, wherein $d$ is the radius of isotropic sphere of the cell. Fig.~\ref{Figure3} illustrates the beamforming gain approximation.

The antenna gain is essential for our analysis on the cell discovery. More specifically, narrowing beam leads to relative higher beamforming gain which eventually increases experienced signal-to-noise ratio (SNR) at Rx. Correspondingly, less transmission duration is required for delivering the same amount of cell discovery related information. This duration, which we referred to as \textit{beam duration}, is a key metric for the mm-wave cell discovery performance evaluation. More details are elaborated in Section III.

APs that are equipped with multiple RF chains perform spatial multiplexing where multiple simultaneous beams can be exploited for broadcast signaling transmission. In this situation, main lobe of the desired beam could be interfered with side lobe of other beams transmitted in parallel when delivering the identical information, and correspondingly the main lobe suffers from inter-beam interference. The side lobe gain is acquired from realistic $16\times8$ uniform rectangular antenna arrays. 

\subsection{Frame Structure and Beam Scan Model}
\begin{figure}[htb]
	\centering
	\includegraphics [scale=0.8]{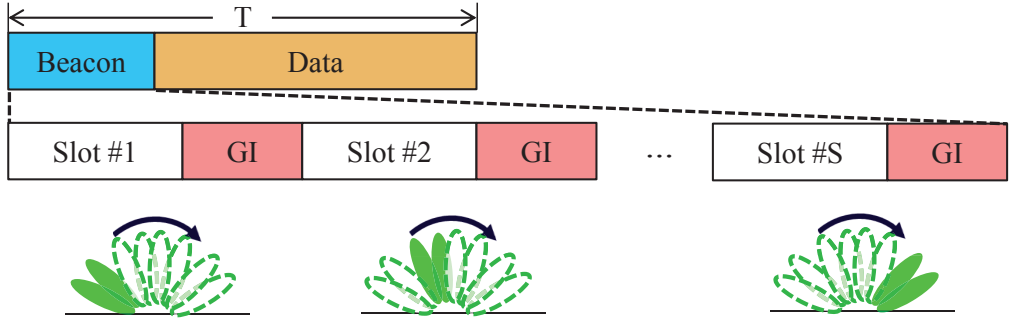}
	\caption{Frame structure including beam scan interval.}\label{Figure4}
\end{figure}

Similar to 802.11ad~\cite{802.11ad} and 802.15.3c~\cite{802.15.3c}, we consider a frame structure consisting of a beacon interval and a data transmission interval, as illustrated in Fig.~\ref{Figure4}. In the beacon interval, Tx broadcasts cell discovery related information via beam scanning over different beam slots. The entire cell is covered by $N$ beam scan areas ($N\leq N_{Tx}$), where AP forms $M$ beams ($1<M\leq N$, i.\,e., limited by the number of RF chains at the Tx) to successively scan these areas. Within the duration of each slot, the formed beams deliver information to UEs located in the corresponding areas. It is obvious that the number of slots is obtained by $S=\frac{N}{M}$. Note that these slots are separated by guard intervals (GIs) reserved for beam switching in the case of hybrid or analog beamforming~\cite{JLuo}. The AP periodically scans the cell via angular search in the beacon interval within each frame, and keep the scanning beam sequence in each frame, namely the beam scan the same area of the cell during the same slot of each frame.

As shown in Fig.~\ref{Figure4} as an example, the entire cell area is covered by eight beam scan areas, where within slot $1$, two Tx beams are scanning the two flat oval-shaped green-filled areas (leftmost), and the duration of the scan is the beam duration mentioned above. Then, the beam rotates and scans the subsequent areas clockwise, two-by-two, with the same duration in each two adjacent beam scan areas. Obviously in the illustrated example, the beacon interval is partitioned into four slots to cover the entire cell (namely $S=4$), and within each slots, two beams are formed to convey the cell discovery related information to UEs in the corresponding areas\footnote{In the illustrated example, the cell is partitioned in $N=N_M\times N_V\, (8=8\times 1)$ scan areas where the beams scan over these area only in horizontal direction. However, the  beam scan areas can be partitioned as $8=4\times 2$ where four areas are close to the AP and the remaining four are far from the AP (with larger elevation angle of antenna). In this case, the beams scans over both vertical and horizontal directions (e.\,g., near to far in vertical direction under same azimuth angle, then clockwise in horizontal direction).}. It is worth noting that multiple simultaneous beams are explicitly exploited in the example. In case single beam exhaustive scan is applied, the beacon interval would be partitioned into eight slots wherein UEs located in each scan area is covered by a single beam during the corresponding slot.

\section{Broadcast Signaling Schemes Design and Analysis}
Similar to~\cite{Alkhateeb, Barati, YingzheLi}, we investigate beamformed cell discovery for mm-wave initial access and specifically focus on broadcast signaling design. Apart from single beam exhaustive scan, multi-beam simultaneous scan are applied by broadcast signaling to exploit the potential benefit of beamformed cell discovery. The notations and system parameters used for the analysis in this section are given in Table~\ref{Table2}.

\begin{table}[htb]
	\centering
	\caption{Broadcast Signaling Schemes Analysis Parameters}
	\label{Table2}
	\resizebox{400pt}{!}{%
	\begin{tabular}{|l|l|l|}
		\hline \\[-1em]
		Notation & Description & Value \\ \hline \\[-1em]
		$P$ & Total Tx power & $30$ dBm \\ \hline \\[-1em]
		$\eta$ & Thermal noise power density & $-174$ dBm/Hz \\ \hline \\[-1em]
		$M_\text{dB}$ & Multiplexing gain in dB scale & $10\log_{10}{M}$ \\ \hline \\[-1em]
		SG & Spreading gain of CD in dB scale  & $10\log_{10}{M}$ \\ \hline \\[-1em]
		$\text{SNR}_\text{TD}$, $\text{SNR}_\text{FD}$, $\text{SNR}_\text{CD}$, $\text{SNR}_\text{SD}$ & Receive SNR of schemes TD, FD, CD, and SD &  \\ \hline \\[-1em]
		$I_\text{B}$ & Inter-beam interference &  \\ \hline \\[-1em]
		$t$ & Beam duration &  \\ \hline \\[-1em]
		$t_\text{GI}$ & Guard interval duration &  \\ \hline \\[-1em]
		$\text{t}_\text{TD}$, $\text{t}_\text{FD}$, $\text{t}_\text{CD}$, $\text{t}_\text{SD}$ & Beam duration of schemes TD, FD, CD, and SD &  \\ \hline \\[-1em]
		$U$ & Amount of cell discovery related information & $128$ bit \\ \hline \\[-1em]
		$n$ & Finite blocklength & $1000$  \\ \hline \\[-1em]
		$\epsilon$ & Block error rate & $0.001$  \\ \hline \\[-1em]
		$C$ & Capacity of AWGN channel &  \\ \hline \\[-1em]
		$V$ & Channel dispersion &  \\ \hline \\[-1em]
		$Q(.)$ & Complementary Gaussian cumulative distribution function &  \\ \hline
	\end{tabular}}
\end{table}

\subsection{Schemes for AP to Broadcast Cell Discovery Information}

To harvest the multiplexing diversity, different broadcast signaling schemes are investigated to enable the AP to broadcast the cell discovery related information. The schemes considered in this paper are described as follows:

1) Time-division (TD): AP scans over the entire cell area with single beam (i.\,e., $1<N\leq N_\text{Tx}$, $M=1$). Here, we refer this scheme as the baseline design, which is especially suitable for Tx with a single RF chain.

2) Frequency-division (FD): If multiple RF chains are available, more than one beams can be formed simultaneously ($1<N\leq N_\text{Tx}$, $1<M\leq N$). In addition to the baseline, this scheme exploits multiple simultaneous beams multiplexed in frequencies. Here, we assume total available bandwidth and transmission power are equally allocated on each beam.

3) Code-division (CD): Similar to FD, in this scheme multiple simultaneous beams are exploited, while multiplexed in codes ($1<N\leq N_\text{Tx}$, $1<M\leq N$). We apply orthogonal codes on the multiplexed beams where a spreading gain SG is expected. The performance of CD with orthogonal codes are discussed in Section III-B and the reason of selecting orthogonal codes are addressed in Appendix A. In addition, total transmission power is also equally allocated on each beam.

4) Space-division (SD): A general scheme, or an extension to TD, is proposed here as SD ($1<N\leq N_\text{Tx}$, $1<M\leq N$). In this scheme, multiple simultaneous beams are formed where no specific multiplexing scheme is addressed. As mentioned in Section II-C, main lobe of the desired beam suffers from inter-beam interference referred to as $I_\text{B}$.

A summary of these schemes is provided in Table~\ref{Table3}.

\begin{table}[htb]
	\centering
	\caption{Broadcast Signaling Schemes}
	\label{Table3}
	\begin{tabular}{|l|l|l|l|}
		\hline \\[-1em]
		Scheme & Multiplexing & $N$ & $M$ \\ \hline \\[-1em]
		TD & Time & $1, 2, 4, ..., 128$ & $1$ \\ \hline \\[-1em]
		FD & Space, time, frequency & $1, 2, 4, ..., 128$ & Divisor of $N$ \\ \hline \\[-1em]
		CD & Space, time, code & $1, 2, 4, ..., 128$ & Divisor of $N$ \\ \hline \\[-1em]
		SD & Space, time & $1, 2, 4, ..., 128$ & Divisor of $N$ \\ \hline
	\end{tabular}
\end{table}

\subsection{SNR and Beam Duration Analysis for Broadcast Signaling Schemes}

The receive SNR at a UE for the broadcast signaling schemes, based on their characteristics summarized in Table~\ref{Table3}, are derived as follows.

For the baseline scheme, the receive SNR is derived as
\begin{equation} 
\text{SNR}_\text{TD} = P+G-L-\eta, \label{eqnTD}
\end{equation}
where $P$, $G$, $L$, and $\eta$ represent the transmission power of the scanning beam, the antenna gain, the path loss, and the thermal noise power density, respectively, all in dB scale. 

Assuming $M$ simultaneous beams are employed, the receive SNR per beam for FD and CD can be written as
\begin{align*}
\text{SNR}_\text{FD} &= P-M_{\text{dB}}+G-L-(\eta-M_{\text{dB}}) \numberthis \label{eqnFD},\\
\text{SNR}_\text{CD} &= P-M_{\text{dB}}+\text{SG}+G-L-\eta \numberthis \label{eqnCD},
\end{align*} 
where $P-M_\text{dB}$ indicates the power split on each beam transmitting $M$ beams simultaneously by FD or CD, and $\eta-M_\text{dB}$ is the noise power density in $\frac{1}{M}$-th bandwidth (frequency band per beam in FD). When applying orthogonal codes for CD, the spreading gain equals to the number of orthogonal codes~\cite{Yang3GCDMA}, namely $\text{SG}=M_{\text{dB}}$. Therefore in terms of SNR, FD and CD perform exactly the same as TD, i.\,e.,
\begin{equation} 
\text{SNR}_\text{TD} = \text{SNR}_\text{FD} = \text{SNR}_\text{CD}.
\label{eqnTDFDCD}
\end{equation}

Similarly, the receive SNR for SD can be derived as
\begin{align*}
\text{SNR}_\text{SD} &= P-M_{\text{dB}}+G-I_{\text{B}}-L-\eta \\
&\leq \text{SNR}_\text{TD}, \numberthis \label{eqnPD}
\end{align*}
where $I_{\text{B}}$ refers to the inter-beam interference.

As mentioned in Section II-C and II-D, beam duration is defined as the duration required for AP to deliver certain amount of information $U$ to UE for cell discovery (e.\,g., cell ID, beam ID, etc.\footnote{In this analysis, the cell detection is also considered as 1-bit information indicating the existence of a cell. Further, frequency and time synchronization requirements has not been treated and the incorporation of the requirements can be put on top of our result and remain as future work.}). It also represents the length of each beam slot in the frame structure illustrated in Fig.~\ref{Figure4}. During this period of time, one or multiple beams are formed by the AP to broadcast the cell discovery related information, where each beam, corresponds to one beam scan area (see Fig.~\ref{Figure4}), is maintained for such long duration to transmit the information. In the next slots, the next group of beams (one or multiple) are formed by the AP to broadcast the information to the UEs in their corresponding beam scan area, and also last the same duration. 

The beam duration, denoted as $t$, is derived according to the achievable rate of AWGN channel with finite blocklength $n$, block error rate $\epsilon$, and bandwidth $B$ as follows~\cite{Polyanskiy}: 
\begin{equation} 
t = \frac {U} { B \Big(C-\sqrt{\frac{V}{n}}Q^{-1}(\epsilon)+\frac{1}{2n}\log_2{n} \Big)},\label{eqnt}
\end{equation}
where $C$ indicates the channel capacity and $V$ is referred to as the channel dispersion that measures the stochastic variability of the channel relative to a deterministic channel with the same capacity, respectively. The two parameters are calculated as
\begin{align*}
C &= \frac{1}{2} \log_2{(1+\text{SNR})}, \numberthis \label{eqnC}\\
V &= \frac{\text{SNR}}{2} \frac{\text{SNR}+2}{(\text{SNR}+1)^2} \log_2^2{e}. \numberthis \label{eqnV}
\end{align*}
Further, $Q(.)$ is the complementary Gaussian cumulative distribution function. 

Due to reduced bandwidth/increased code length of FD/CD multiplexed on $M$ simultaneous beams, beam durations of FD and CD, denoted as $t_\text{FD}$ and $t_\text{CD}$, respectively, can be written as
\begin{equation} 
t_\text{FD} = t_\text{CD} = M t_\text{TD}, \label{eqntTDFD}
\end{equation}
where $t_\text{TD}$ represents the beam duration of TD. Similarly, beam duration of SD, $t_\text{SD}$, should be
\begin{equation} 
t_\text{SD} \geq t_\text{TD}. \label{eqntTDPD}
\end{equation}

It can be observed from~\eqref{eqntTDFD} and~\eqref{eqntTDPD} that the four schemes in Table~\ref{Table3} could lead to different beam durations. In the rest of this paper, we develop
and verify a general analytical framework that can quantify the impact of various broadcast signaling schemes and their corresponding beam durations on the mm-wave cell discovery performance.

\section{Cell Discovery Analysis and Performance Metrics}
The objective of this section is to characterize the performance of mm-wave cell discovery taking into account the impact of the four broadcast signaling schemes. The performance metrics, namely the CDL and the OH, are introduced in Section IV-A. In Section IV-B, we develop an analytical framework with knowledge of beacon timing between AP and UE for initial access to characterize the introduced performance metrics. In addition to this framework, we address analysis studies on framework without knowledge of beacon timing and consider general beacon interval integration, in Section IV-C and Section IV-D respectively, as two extension frameworks. Under the framework without knowledge of beacon timing, the analysis reveals succinct characterizations of the CDL and OH.

\subsection{Performance Metrics}
Cell discovery is a basic prerequisite to any communication and is an essential component of initial access procedure. For a UE to detect the presence of an AP, the cell periodically broadcasts discovery related information via angular search in the beacon interval of the frame, introduced in Section II-D and illustrated in Fig.~\ref{Figure4}. 
When the AP forms the beam and starts to scan the area in which a UE is located, the UE should already be active in order to receive the complete cell discovery related information. Otherwise, it has to wait till the end of the complete beam scan phase (the beacon interval in this frame, where the AP scans over the entire cell area), plus the potential duration from AP starting the next round of scan to the beam reaching its corresponding area, and a complete beam duration. It is worth noting that the above situation is only valid in case UE successfully decode the information with the probability $1-\epsilon$, otherwise it has to wait rounds of frames till successful discovering the cell (geometric distribution of successful discovery).

The above additive time, which accounts for the delay a UE experiences from the time stamp it transits from idle to active ($z_\text{active}$), to the time stamp when it receive and decode the complete cell discovery information, is referred to as the CDL. We further define the OH as the portion of the beacon interval in one frame to depict the burden of the AP to inform its presence to its attached UEs. It is clear that the CDL strongly depends on $z_\text{active}$ and the beam scan area the UE is located in. Moreover, the selection of $N$ and $M$, as well as the broadcast signaling schemes, have significant impacts on the OH.

Without loss of generality, we assume $z_\text{active}$ is uniformly distributed in the entire frame (not all UEs are necessarily active at the same frame), which corresponds to the realistic scenario that UE could be active at anytime. UE keeps active till the end of a successful cell discovery. We further define the time stamp of which the beams start to scan the UE located area as $z_\text{scan}$. It is intuitively that the smallest CDL refers to the case that when $z_\text{active}=z_\text{scan}$, where the latency is just the beam duration $t$. On the contrary, the largest CDL refers to the situation that a UE fails to detect the cell in $K$ frames (with the probability $\epsilon^K\rightarrow0$, when $K\rightarrow\infty$), and the latency approaches infinity. Here, $K\rightarrow\infty$ indicates the fact that all UEs eventually succeed in discovering the presence of the cell.

In the following subsections, we provide precise analytical frameworks to study the more general case, such that UE is active arbitrarily in a frame and there is no specific relation between $z_\text{active}$ and $z_\text{scan}$. We are interested in which factors are key to characterize the average CDL and the OH. In Section IV-B and IV-C, a complete beacon interval in integrated into one frame. Extensions, such as long frame containing multiple beacon intervals, OH-limited case where a normal frame containing a beacon interval is followed by several consecutive data-only frames, and the case where a complete beacon is equally separated into several consecutive frames, are studied in Section IV-D.

\subsection{Cell Discovery with Knowledge of Beacon Timing and Single Complete Integrated Beacon Interval}
The beacon timing between AP and UE could be achieved by the well-know control plane (C-plane) and user plane (U-plane) split, where the timing is obtained via a macro cell operating at legacy bands, so that all the UEs are instructed to listen to the mm-wave AP beacon and aligned to the beginning of a frame. 
As UE is uniformly ``dropped'' in the cell and the cell is covered by beams scan with $S$ slots (each slot corresponds to a scan area which is depicted in Fig.~\ref{Figure4}, and $S$ is the number of the beam slots summarized in Table~\ref{Table1}), the probability of the event ``UE is located in the area corresponding to slot $j$'', denoted as $p_\mathrm{O}$, can be written as $p_\mathrm{O}=\frac{1}{S}$. For simplicity, in the rest of the paper we represent the event ``UE is located in the area corresponding to slot $j$'' as a short term ``located in slot $j$''. 

For this event, UE attempts to decode the delivered information with error rate $\epsilon$. In this case, the latency conditioned to the event, denoted as $t_\mathrm{O}$, is given as follows:
\begin{align*}
t_\mathrm{O} = &\; j \text{ beam slots} \cdot \text{probability of success}
  \\
& + (\text{one frame} + j \text{ beam slots}) (1 - \text{probability of success}) \cdot \text{probability of success}\\
& + ... \\
& + (K{-}1\text{ frames} + j \text{ beam slots}) (1 - \text{probability of success})^{K{-}1} \cdot \text{probability of success} \\
& + K\text{ frames} \cdot (1 - \text{probability of success})^{K}. \numberthis \label{eqntO}
\end{align*}
In~\eqref{eqntO}, $j$ beam slots includes $j-1$ beam slots and $j-1$ GI, plus a extra beam slot $j$, which is the decoding time. For simplicity, we denote $t+t_\text{GI}$ as $t'$ and use the shorthand denotation in the rest of the paper. Then the average CDL over 
the uniform distribution of $j$ from $1$ to $S$ for the cell discovery with knowledge of beacon timing, $\overline{T}_\text{w}$, can be derived as
\begin{align*}
\overline{T}_\text{w}
&= \sum_{j=1}^{S} p_\mathrm{O} t_\mathrm{O} \\
&= \sum_{j=1}^{S} \frac{1}{S} \Bigg( \sum_{k=0}^{K-1} \big(k T + (j-1)t' + t \big) \epsilon^k(1-\epsilon) + K T \epsilon^K \Bigg) \\
&= (1-\epsilon^K)(\frac{S+1}{2} t + \frac{S-1}{2} t_\text{GI} + \frac{\epsilon}{1-\epsilon} T). \numberthis \label{eqn0}
\end{align*} 
When $K\rightarrow \infty$, we have
\begin{equation} 
\overline{T}_\text{w} = \Big(\frac{S+1}{2} t + \frac{S-1}{2} t_\text{GI} + \frac{\epsilon}{1-\epsilon} T\Big).
\label{eqn00}
\end{equation}

In short,~\eqref{eqn00} is explained by noticing that if there is a decoding error, with the probability $\epsilon$, then the discovery procedure fails, and the UE must wait a whole round and try again, assuming that all errors can be revealed.

\subsection{Cell Discovery without Knowledge of Beacon Timing and Single Complete Integrated Beacon Interval}
In fact, it is not viable to get very accurate time and frequency synchronization from legacy bands for mm-wave, as the time domain and frequency domain experience very different granularities~\cite{D42}. Accordingly, only rough synchronization is expected to be achieved in legacy bands assisted networks. In addition, the request of cell discovery for a UE emerges randomly in practice, and correspondingly, the CDL should be counted exactly from the emergence of this request to the moment of successful discovery. Therefore, in this subsection we address the cell discovery without knowledge of beacon timing by considering the UE active time stamp $z_\text{active}$.

\begin{figure}[htb]
	\centering
	\includegraphics [scale=0.8]{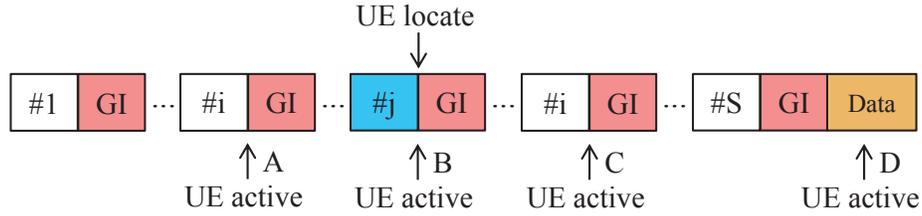}
	\caption{Illustration of the four cell discovery scenarios.}\label{FigureS}
\end{figure}

Mapping UE located area to the beam slot, there are four scenarios distinguished by $z_\text{active}$ and slot $j$ (the scan area corresponding to slot $j$), illustrated by Fig.~\ref{FigureS}. Without loss of generality, we assume UE is active at slot $i$. For scenarios A to C, $z_\text{active}$ is within the beacon interval, while in the last scenario, UE is active in the data interval. Besides, for scenario A, UE has a chance to finish the cell discovery procedure in its active frame (we referred to as ``the current frame'') in case successfully decoding the information, as it is active before beam scans its located area. For the remaining scenarios, UE waits at least one round and try to detect the cell in the next frames. In the following, we will describe the four scenarios and analyze their specific average CDLs in details. The corresponding OHs of are derived afterwards.

\subsubsection*{Scenario A -- UE Becomes Active before Beam Scans Its Location}~\\
\indent In this scenario we have $j>i$, therefore UE can ``catch'' the beam to decode the information when the beam is scanning its location in the current frame. The probability of scenario A, $p_\mathrm{A}(i)$, which is the event ``UE is active in slot $i$ conditioned to UE locates in slot $j$ from $i+1$ to $S$'', is written as
\begin{align*}
p_\mathrm{A}(i) &= p(\text{active in slot $i$}) p(\text{located in slot $j$}) \\
&=\frac{t'}{T} \frac{S-i}{S}. \numberthis \label{eqn1}
\end{align*} 
Here, as $z_\text{active}$ is uniformly distributed in the entire frame, the probability of UE active at slot $i$ is $\frac{t'}{T}$. Similarly, due to the uniform distribution of UE location in the entire cell, the probability of UE locating in beam slot $j$ is $\frac{S-i}{S}$.

For this event, UE attempts to decode the delivered information with error rate $\epsilon$ in the current frame. The latency conditioned to the event, denoted as $t_A(i)$, consists of three parts:
\begin{itemize}
	\item Latency in slot $i$: $\int_{(i-1)t'}^{it'} \! \frac{ it'-z_\text{active} }{it'-(i-1)t'}  \, \mathrm{d}z_\text{active}=\frac{t'}{2}$.
	\item Latency between the end of slot $i$ and the beginning of slot $j$: $({j}{-}{i}{-}{1})t'$.
	\item Beam decoding latency: $t$.
\end{itemize}

In case UE is not able to successfully decode the information in the current frame, it has to wait at least till beam scans its location in the next frame. In summary, $t_\mathrm{A}(i)$, averaging over the uniform probability of $j$ from $i+1$ to $S$, can be derived as
\begin{align*}
t_\mathrm{A}(i)
&= \bigg( \frac{t'}{2} + \frac{\sum_{j=i+1}^{S}(j-i-1)t'}{S-i}  + t \bigg) (1-\epsilon) + \sum_{k=1}^{K-1} \bigg( \frac{t'}{2} + T - it' + (k-1)T + \frac{\sum_{j=i}^{S-1}jt'}{S-i} + t \bigg) \\
&\quad\quad \cdot \epsilon^k(1-\epsilon) + \bigg( \frac{t'}{2} + T - it' + (K-1)T \bigg) \epsilon^K. \numberthis \label{eqn2}
\end{align*} 

In~\eqref{eqn2}, the term $\frac{t'}{2}+T-i({t}{+}t_\text{GI})$ indicates the latency of the current frame. The term $\frac{\sum_{j=i}^{S-1}jt'}{S-i}{+}{t}$ indicates the latency of successful decoding frame. 

It is assumed that in this scenario UE is active before its located slot, which means the last active slot should be slot $S-1$. Thus, the average CDL for scenario A, denoted as $\overline{T}_\mathrm{A}$, is distributed over the uniform probability of $i$ from $1$ to $S-1$, and can be derived as
\begin{equation} 
\overline{T}_\mathrm{A} = \sum_{i=1}^{S-1} p_\mathrm{A}(i) t_\mathrm{A}(i).
\label{eqn3}
\end{equation} 

The detailed derivations of the average CDL for scenario A, as well as for the remaining scenarios, are shown in Appendix B.

\subsubsection*{Scenario B -- UE Is Active when Beam Is Scanning Its location}~\\
\indent In this scenario we have $j=i$. The probability of scenario B, $p_\mathrm{B}(i)$, which is the event ``UE is active in slot $i$ conditioned to UE locates in slot $j=i$'', is written as
\begin{align*}
p_\mathrm{B}(i) &= p(\text{active in slot $i$}) p(\text{located in slot $j$}) \\
&=\frac{t'}{T} \frac{1}{S}. \numberthis \label{eqn4}
\end{align*} 

Here, UE fails in cell discovery within the current frame, as it is not able to decode the beacon information within a complete beam duration (we assume this can be only done when UE is active at exactly the beginning of the scanning slot\footnote{Just for the case that UE experiences high SNR and can decode even from just a part of the receive signal.}). Therefore, UE attempts to decode the delivered information with error rate $\epsilon$ in the next frames. The latency conditioned to the event, denoted as $t_\mathrm{B}(i)$, consists of the following parts:
\begin{itemize}
	\item Latency in slot $i$: $\frac{t'}{2}$.
	\item Latency till end of the current frame: $T-it'$.
	\item Latency in the next frame: $(j-1)t'+t$.
\end{itemize}

In case UE is not able to successfully decode the information in the next frame, it has to wait another whole frame. In summary, $t_\mathrm{B}(i)$ can be derived as
\begin{align*}
t_\mathrm{B}(i)
&= \frac{t'}{2} + T-it' + \sum_{k=0}^{K-2} \big( kT + t + (i-1)t' \big) \epsilon^k(1-\epsilon) + (K-1)T\epsilon^{K-1}. \numberthis \label{eqn5}
\end{align*} 

As UE is not able to finish the cell discovery in the current frame, its discovery trials are counted from the second ($k=0$) frame to the last ($k=K-1$), which means no matter UE succeeds in which frame, the latency $\frac{t'}{2}+T-i({t}{+}t_\text{GI})$ in the current frame needs to be included in the $t_\mathrm{B}(i)$. 

The average CDL over the uniform probability of $i$ from $1$ to $S$ for scenario B, denoted as $\overline{T}_\mathrm{B}$, can be derived as
\begin{equation} 
\overline{T}_\mathrm{B} = \sum_{i=1}^{S} p_\mathrm{B}(i) t_\mathrm{B}(i).
\label{eqn6}
\end{equation}

\subsubsection*{Scenario C -- UE Is Active after Beam Has Scanned Its Location}~\\
\indent In this scenario we have $j<i$. The probability of scenario C, $p_\mathrm{C}(i)$, which is the event ``UE is active in slot $i$ conditioned to UE locates in slot $j$ from $1$ to $i-1$'', is written as
\begin{align*}
p_\mathrm{C}(i) &= p(\text{active in slot $i$}) p(\text{located in slot $j$}) \\
&=\frac{t'}{T} \frac{i-1}{S}. \numberthis \label{eqn7}
\end{align*}  

Similar to scenario B, UE fails in the cell discovery within the current frame, and attempts to decode the delivered information with error rate $\epsilon$ in next frames. The latency conditioned to the event, denoted as $t_\mathrm{C}(i)$, are same as that of scenario B. In case UE is not able to successfully decode the information in the next frame, it has to wait another whole frame. In summary, $t_\mathrm{C}(i)$ averaging over the uniform probability of $j$ from $1$ to $i-1$ can be derived as
\begin{align*}
t_\mathrm{C}(i)
&= \frac{t'}{2} + T-it' + (K-1)T\epsilon^{K-1} + \sum_{k=0}^{K-2} \bigg( kT + t + \frac{\sum_{j=1}^{i-1} (j-1)}{i-1}t' \bigg)\epsilon^k(1-\epsilon). \numberthis \label{eqn8}
\end{align*} 

Accordingly, the average CDL over the uniform probability of $i$ from $2$ to $S$ for scenario C, denoted as $\overline{T}_\mathrm{C}$, can be derived as
\begin{equation} 
\overline{T}_\mathrm{C} = \sum_{i=2}^{S} p_\mathrm{C}(i) t_\mathrm{C}(i).
\label{eqn9}
\end{equation} 
Note that it is assumed that in this scenario UE is active after it located slot, which means the first active slot should be slot $2$.

\subsubsection*{Scenario D -- UE Is Active in Data Interval}~\\
\indent It is clear that in this scenario, UE can complete the cell discovery procedure with error rate $\epsilon$ only in the next frames. The probability of scenario D, $p_\mathrm{D}(i)$, which is the event ``UE is active in data interval of the current frame'', is written as
\begin{equation}
p(\text{active in data interval}) =\frac{T-St'}{T}. \numberthis \label{eqn10}
\end{equation}

The latency conditioned to the event, denoted as $t_\mathrm{D}(i)$, consists of the following parts:
\begin{itemize}
	\item Latency in the current frame: $\int_{St'}^{T} \! \frac{ T-z_\text{active} }{T-St'}  \, \mathrm{d}z_\text{active}$=$\frac{T-St'}{2}$.
	\item Latency in the next frame: $({j}{-}{1})t'+t$.
\end{itemize}

Then, the average CDL over the uniform probability of $j$ from $1$ to $S$ for scenario D, denoted as $\overline{T}_\mathrm{D}$, can be derived as
\begin{align*}
\overline{T}_\mathrm{D}
&= \frac{T-St'}{T} \Bigg( \frac{T-St'}{2}  + (K-1)T\epsilon^{K-1} + \sum_{k=0}^{K-2} \bigg( kT + t + \frac{\sum_{j=1}^{S} (j-1)}{S}t' \bigg) \epsilon^k(1-\epsilon) \Bigg).  \numberthis \label{eqn11}
\end{align*} 

\subsubsection*{The Overall Average CDL}~\\
\indent Combining the four scenarios, the average latency for cell discovery without knowledge of beacon timing and single complete integrated beacon interval, $\overline{T}_\text{w/o}$, can be derived as
\begin{equation} 
\overline{T}_\text{w/o} = \overline{T}_\mathrm{A} + \overline{T}_\mathrm{B} + \overline{T}_\mathrm{C} + \overline{T}_\mathrm{D}.
\label{eqn13}
\end{equation}
When $K\rightarrow \infty$, we have
\begin{equation} 
\overline{T}_\text{w/o} = t + \frac{1+e}{2(1-e)} T.\label{eqn14}
\end{equation}

This result is quite inspiring as it shows that the average CDL depends only on beam duration, error rate, and frame length, and is irrespective of the option of GI. We consequently argue that analog/hybrid beamforming performs as well as digital beamforming in the context of CDL, if the entire beacon interval can be accommodated in one frame. However, long GI of hybrid beamforming definitely leads to high signaling OH. 

Considering the analysis of different broadcast signaling schemes patterns in Section III, the average CDL of TD, FD, CD, and SD, denoted as $\overline{T}_\text{TD}$, $\overline{T}_\text{FD}$, $\overline{T}_\text{CD}$, and $\overline{T}_\text{SD}$, respectively, can be written as
\begin{equation} 
\overline{T}_\text{TD} = t_\text{TD} + \frac{1+e}{2(1-e)} T,\label{eqnLatencyTD}
\end{equation}
\begin{align*}
\overline{T}_\text{FD} = \overline{T}_\text{CD} &= M \text{t}_\text{TD}+ \frac{1+e}{2(1-e)}T \\
&\geq \overline{T}_\text{TD},\numberthis \label{eqnLatencyFDCD}
\end{align*}
and
\begin{align*}
\overline{T}_\text{FD} (=\overline{T}_\text{CD}) \geq  \overline{T}_\text{SD} &= \text{t}_\text{SD}+ \frac{1+e}{2(1-e)}T \\
&\geq \overline{T}_\text{TD}.\numberthis \label{eqnLatencySD}
\end{align*}

Similarly, for OH, we have
\begin{align*}
\text{OH}_\text{TD} &= \frac{(t_\text{TD}+t_\text{GI}) S_{\text{TD}}}{T} = \frac{N(t_\text{TD}+t_\text{GI})}{T}, \numberthis \label{eqnOHTD}
\end{align*} 

\begin{align*}
\text{OH}_\text{FD} = \text{OH}_\text{\text{CD}} &= \frac{(t_\text{FD/CD}+t_\text{GI}) S_{\text{FD/CD}}}{T} = \frac{\frac{N}{M} (Mt_\text{TD}+t_\text{GI}) }{T} = \frac{Nt_\text{TD} + \frac{N}{M}t_\text{GI} }{T} \\
& \leq \text{OH}_\text{\text{TD}}, \numberthis \label{eqnOHFDCD}
\end{align*} 
and
\begin{align*}
\text{OH}_\text{SD} &= \frac{(t_\text{SD}+t_\text{GI}) S_{\text{SD}}}{T} = \frac{\frac{N}{M} (t_\text{SD}+t_\text{GI}) }{T} \\
&\leq \frac{Nt_\text{TD} + \frac{N}{M}t_\text{GI} }{T} = \text{OH}_\text{FD} (= \text{OH}_\text{\text{CD}}) \\
&\leq \frac{N (t_\text{TD}+t_\text{GI}) }{T} = \text{OH}_\text{\text{TD}}. \numberthis \label{eqnOHSD}
\end{align*}

To summarize, we argue that:

\begin{itemize}
	\item The average CDL depends only on beam duration, error rate, and frame length, and is irrespective of the option of GI.
	\item Single beam exhaustive scan (TD) outperforms all broadcast signaling schemes in terms of the CDL, but results in high OH.
	\item Simultaneous multi-beam scan (FD/CD/SD) can significantly reduce the OH, and provide the flexibility to achieve trade-off between CDL and OH.
\end{itemize}

\subsection{Cell Discovery with General Integrated Beacon Integration}
\subsubsection*{Long Frame Containing Multiple Beacon Intervals}
\begin{figure}[htb]
	\begin{minipage}{\textwidth}
		\centering
		\includegraphics[scale=0.8]{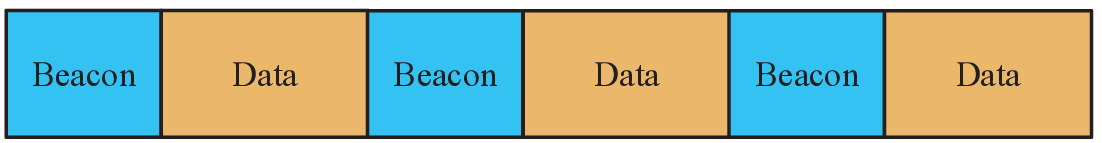}
		\subcaption{Long frame with $W=3$ beacon intervals}
		\label{Figure6:a}\par \medskip \vfill
		\includegraphics[scale=0.8]{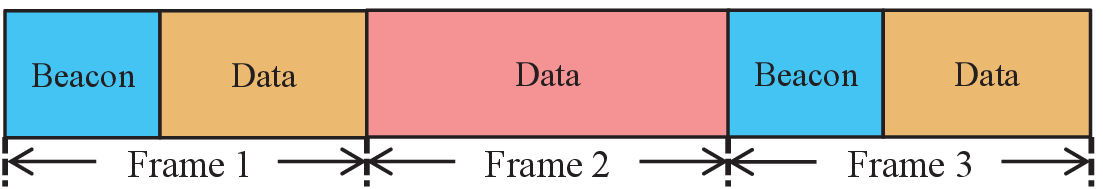}
		\subcaption{Insert beacon interval in every $V=2$ frames}
		\label{Figure6:b}\par \medskip \vfill
		\includegraphics[scale=0.8]{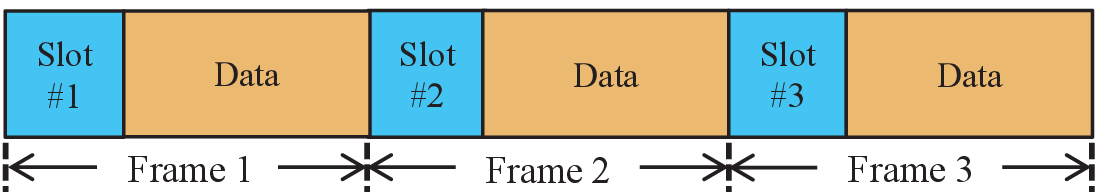}
		\subcaption{A beacon interval is equally separated into $X=3$ frames}
		\label{Figure6:c}
	\end{minipage}
	\caption{General integrated beacon interval design considering (a) long frame with $W=3$ beacon intervals, (b) inserting beacon interval in every $V=2$ frames, and (c) separating beacon interval equally into $X=3$ frames.}\label{Figure6}
\end{figure}

~\\
In the previous sections, only one complete beacon interval is accommodated in each frame. However, considering frame structure design diversity, such as frame structure supporting dynamic time-division duplex (TDD), multiple beacon intervals could be incorporated in one frame due to the requirement of flexible adjustment of DL and UL data. In this case, we enable the frame structure that accommodates $W$ uniformly distributed and separated beacon intervals. An example of the considered frame structure is illustrated in Fig.~\ref{Figure6:a}. Other distributions and/or separation design of beacon intervals can be put on top of this case.

Intuitively, this long frame can be ``partitioned'' into $W$ consecutive subframes where each subframes is mapping to the typical frame studied in Section IV-B and IV-C. More specifically, the original cell discovery supposed to be done in $K$ frames with a single complete integrated beacon interval, is now ``squeezed'' to $\frac{K}{W}$ subframes. Correspondingly, decreased latency $\frac{T}{W}$, which is a span of the frame length $T$, can be expected in the CDL of this case. On the other hand, from the mathematical point of view, the CDL of this case can be derived by substituting $K$ in~\eqref{eqn2},~\eqref{eqn5},~\eqref{eqn8}, and~\eqref{eqn11} with $\frac{K}{W}$, and then the average CDL of this case, denoted as $\overline{T}_\text{long-frame}$, is written as
\begin{equation} 
\overline{T}_\text{long-frame} = t + \frac{1+e}{2W(1-e)} T.\label{eqn20}
\end{equation}
Obviously, the corresponding OH is $W$ times of the original one.

\subsubsection*{OH-Limited Frame without Beacon Intervals}~\\
In some scenarios, large GI, which is reserved for beam switching in the case of hybrid or analog beamforming, would lead to high OH. In such case, incorporating beacon interval in every frame may not fulfill the requirement of fixed or limited OH. Consequently, schemes that alleviate resource waste due to high OH are proposed in 802.11ad~\cite{802.11ad}, including insertion of several data-only frames between two consecutive typical frames, and long beacon split and relocation to sequences of frames. An example of beacon insertion in every $V=2$ frames (namely insert a data-only frame in-between every $V=2$ typical frames) is illustrated in Fig.~\ref{Figure6:b}.

Similarly, in this case each typical frame can be treated as to be ``prolonged'' to a superframe with $V-1$ data-only frames . Therefore, the original cell discovery supposed to be done in $K$ frames with a single complete integrated beacon interval, is now ``equivalent'' to a superframe with the length $VT$. Correspondingly, additional latency $VT$, which is also a span of the frame length $T$, can be expected in the CDL of this case. On the other hand, from the mathematical point of view, the CDL of this case can be derived by substituting $K$ in~\eqref{eqn2},~\eqref{eqn5},~\eqref{eqn8}, and~\eqref{eqn11} with $VK$, and then the CDL, denoted as $\overline{T}_\text{OH-limited}$, is written as
\begin{equation} 
\overline{T}_\text{OH-limited} = t + \frac{1+e}{2(1-e)} VT.\label{eqn21}
\end{equation}
In the context of frame-scale, the OH of this case remains as the original one. However, when considering the OH of a superframe considering both typical and data-only frames, the OH is $\frac{1}{V}$ of the original one.

\subsubsection*{Complete Beacon Interval Separation}~\\
Similar to the previous OH-limited case, the separation of a complete beacon interval into consecutive frames is another way of resource saving proposed in 802.11ad~\cite{802.11ad}. In this case, the beacon interval is equally separated into several consecutive frames and within each frame, the AP scans only partial angular areas (corresponding to beam slots). An example of beacon separation into $X=3$ frames is illustrated in Fig.~\ref{Figure6:c}.

In this case, each frame can be also treated as to be ``prolonged''. However, for a UE in a scanning area, if it miss the beams in its active frame, it needs to wait $X$ frames and then try to decode the delivered information, instead of decoding in the next frame described in Section IV-B and IV-C. Consequently, the term $\frac{1+e}{2(1-e)} T$ in ~\eqref{eqn2}), is $X^2$ times as before, because now between two cell discovery trial there are $X$ frames, compared to a single frame in the situation of Section IV-B and IV-C. Then the CDL, denoted as $\overline{T}_\text{separation}$, is written as
\begin{equation} 
\overline{T}_\text{separation} = Xt + \frac{1+e}{2(1-e)} X^2T.\label{eqn22}
\end{equation}
In the context of frame-scale, the OH of this case is $X$ times less as the original one, however when considering OH of a superframe including the complete beacon interval, the OH remains as the original one.

\section{Numerical Evaluation and Discussion: Broadcast signaling Schemes Impact and Design Insights}
In this section, we evaluate the performance of the proposed broadcast signaling schemes for mm-wave beamformed cell discovery. We also provide some insights into the questions raised in Section I-B. As the baseline scheme is the most straightforward broadcast signaling design which could be potentially implemented by the mm-wave systems~\cite{802.11ad}, the impact of baseline scheme on cell discovery as well as beamforming architecture design is addressed in Section V-A. Then, based on the analyses in Section IV-C and IV-D, we compare the expected CDL and OH for the four broadcast signaling schemes in Section V-B. Finally, we compare the performance of cell discovery versus block error rate in Section V-C, and show how to select the error rate could be beneficial.

The simulation results in this section adopt the system model and broadcast signaling schemes design described in Section II and Section III. The system carrier frequency and bandwidth are $f=28$ GHz and $B=1$ GHz, respectively. One AP is located at the center of the cell, and $100$ UEs are uniformly dropped in the angular domain of the cell.
The adopted propagation and antenna model are explained in Section II-B and II-C. It is worth noting that UEs are assumed
to be almost stationary so the path loss and shadowing values are fixed during the simulation duration. Simulation samples are averaged over 1000 independent snapshots. The default system parameter values are summarized in Table~\ref{Table1} and Table~\ref{Table2}.

\subsection{Impact of the Baseline Broadcast Signaling Scheme on Millimeter Wave Cell Discovery and Beamforming Architecture Design}
\subsubsection*{How wide should the beam be (how to select $N$)?}
\begin{figure}[htb]
	\centering
	\begin{subfigure}{0.48\textwidth}
		\includegraphics[scale=0.55]{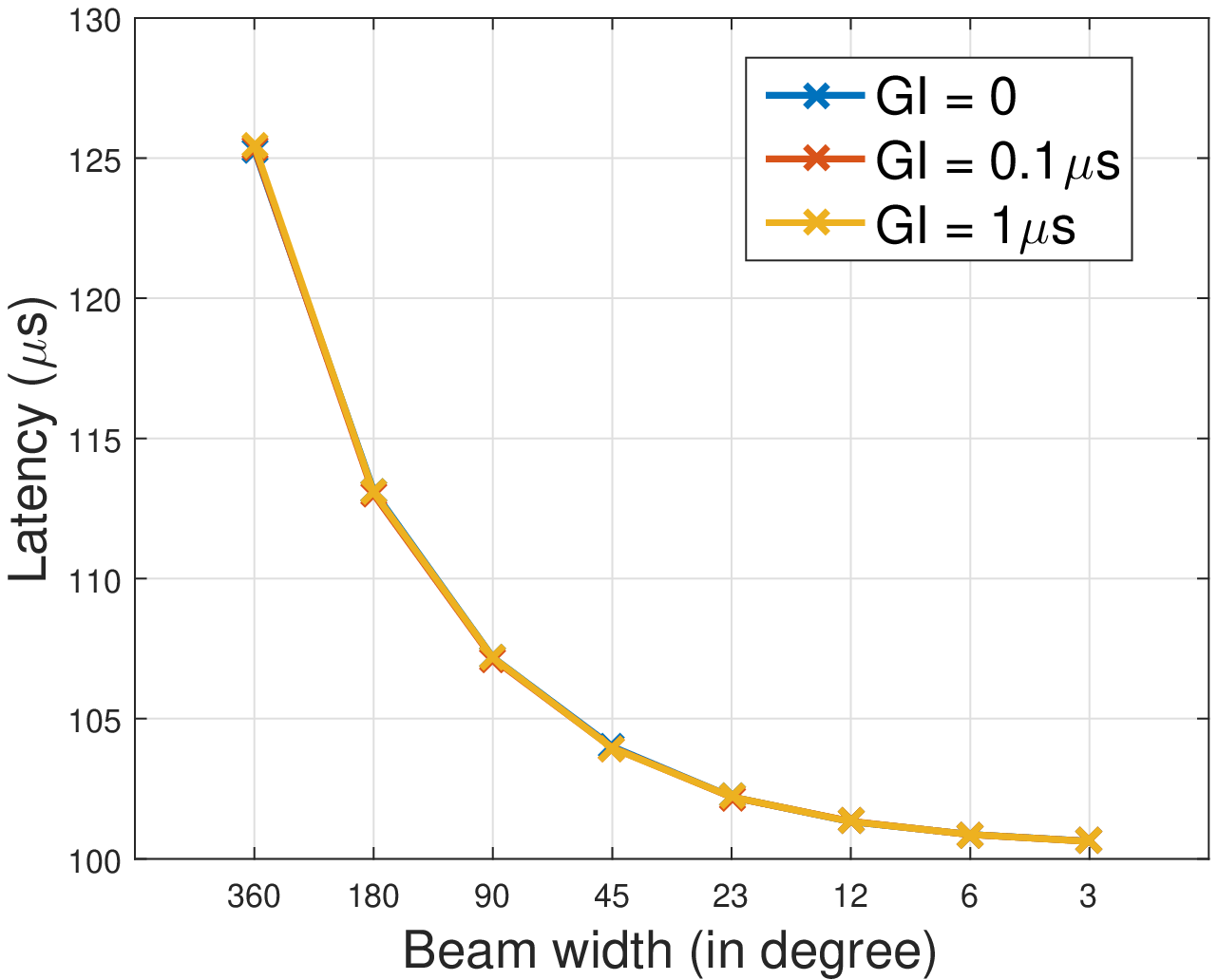}
		\caption{Latency}
		\label{Figure7:a}
	\end{subfigure}
	~ 
	\begin{subfigure}{0.48\textwidth}
		\includegraphics[scale=0.55]{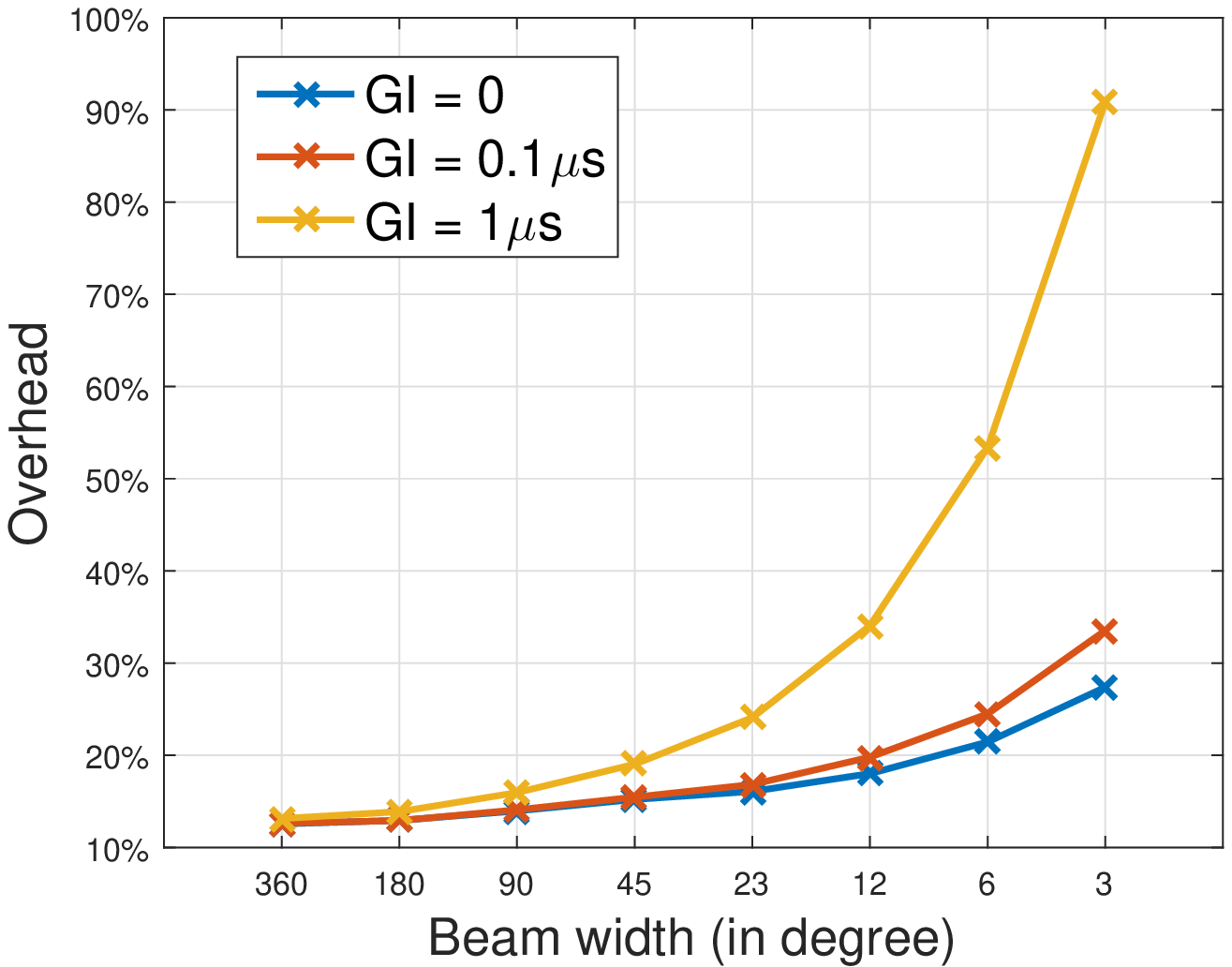}
		\caption{Overhead}
		\label{Figure7:b}
	\end{subfigure}
	\caption{Performance of TD for cell discovery without knowledge of beacon timing and single complete integrated beacon interval are compared in terms of (a) average CDL and (b) OH.}\label{Figure7}
\end{figure}

~\\
In Fig.~\ref{Figure7:a} and Fig.~\ref{Figure7:b}, we plot the average CDL and the OH of TD for cell discovery without knowledge of beacon timing and single complete integrated beacon interval, defined in~\eqref{eqnLatencyTD} and~\eqref{eqnOHTD}, respectively, versus the beam width ($=\frac{360^\circ}{N}$, where $N$ is number of beam scan areas) for different GIs. Here, $\text{GI}=0$ indicates the digital beamforming architecture without beam switch time, and other two values refer to analog/hybrid beamforming with different beam switching times $\text{GI}=0.1$ $\mu$s and $\text{GI}=1$ $\mu$s. Results shows that the analytical result in~\eqref{eqn14} is accurate, where the average CDL is independent of the selection of GI (Tiny fluctuations of the curves refers to non-ideal averaging in the simulations). With thiner beam (larger $N$), the latency enhanced mainly because of the increase in the beamforming gain $G$ in~\eqref{eqnAntennaGain} (and correspondingly decreases the beam duration $t$). Further, we note that the OH degrades when the GI becomes larger, which corresponds to our theoretical analysis that larger GI leads to higher OH.

In summary, these results indicate that analog/hybrid beamforming performs as well as the digital beamforming in terms of the CDL. Furthermore, thiner beam (larger $N$) significantly decrease the CDL. These thiner beams, however, lead to higher OH.

\subsection{Cell Discovery Performance Comparison for Different Broadcast Signaling Scheme}
In this section, based on~\eqref{eqnLatencyTD}--\eqref{eqn21}, we compare the CDL and the OH of the four broadcast signaling schemes in Table~\ref{Table3}. In making the comparison, we consider only the digital beamforming architecture, i.\,e., $\text{GI}=0$, where the performance of the schemes with other GI values can be validated similar to the results. Here, performance evaluation of both cell discovery with single complete integrated beacon interval in Section IV-C and with general integrated beacon interval in Section IV-D are demonstrated.

\subsubsection*{Is it beneficial to exploit multi-beam simultaneous scan?}
\begin{figure}[htb]
	\centering
	\begin{subfigure}{0.48\textwidth}
		\includegraphics[scale=0.55]{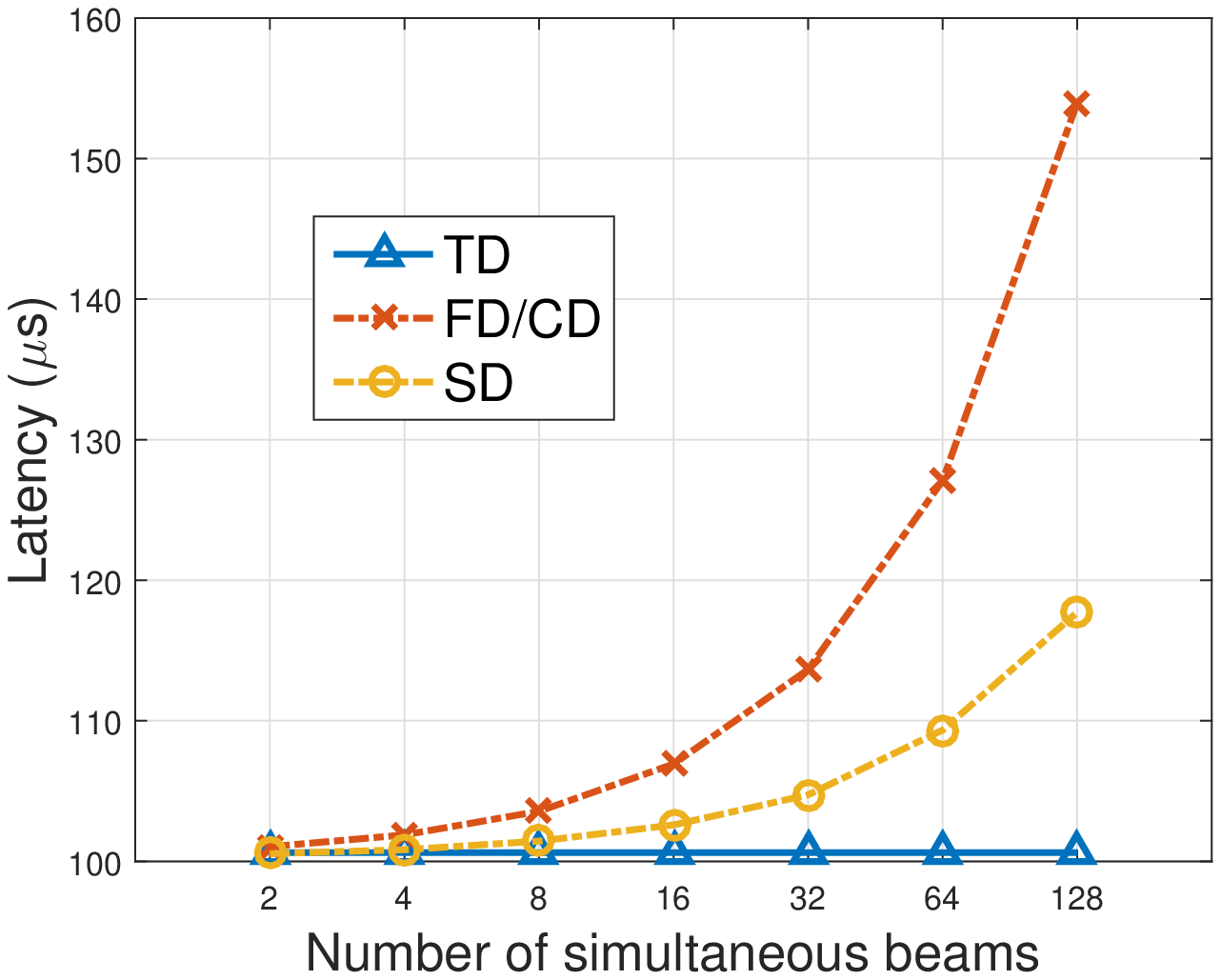}
		\caption{Latency}
		\label{Figure8:a}
	\end{subfigure}
	~ 
	\begin{subfigure}{0.48\textwidth}
		\includegraphics[scale=0.55]{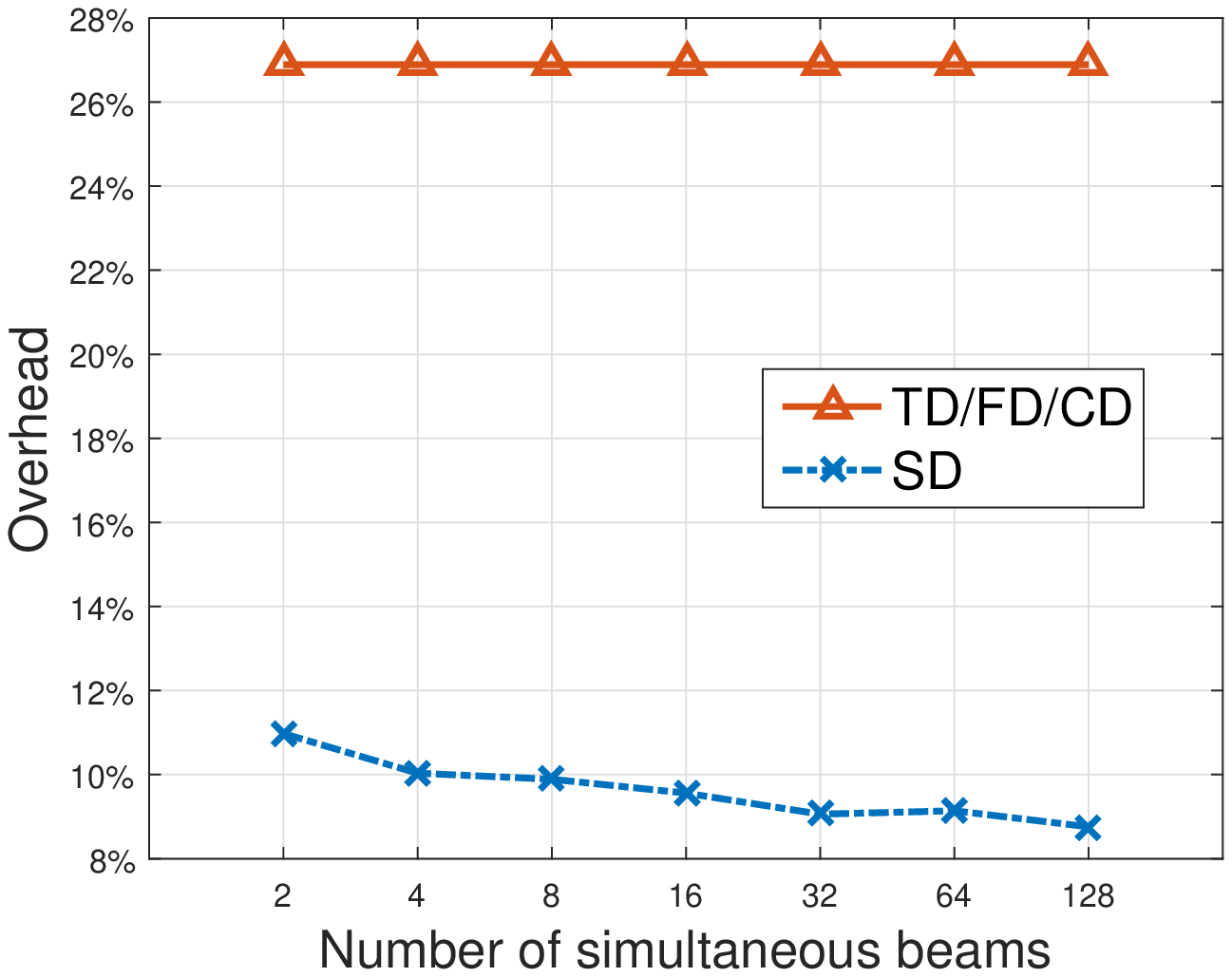}
		\caption{Overhead}
		\label{Figure8:b}
	\end{subfigure}
	\caption{Performance of different broadcast signaling schemes for cell discovery without knowledge of beacon timing and single complete integrated beacon interval are compared in terms of (a) average CDL and (b) OH.}\label{Figure8}
\end{figure}

~\\
To get some insights into the answer of this question, we plot the average CDL and the OH of different broadcast signaling schemes without knowledge of beacon timing and single complete integrated beacon interval, defined in~\eqref{eqnLatencyTD}--\eqref{eqnOHSD}, versus the number of simultaneous beams $M$ in Fig.~\ref{Figure8}. The number of beam scan areas $N$ is assumed to be $128$.

The results in Fig.~\ref{Figure8:a} show that when the number of scan beam areas is fixed, the baseline scheme TD, which is the single beam exhaustive scan, achieves the lowest average CDL. FD and CD performs exactly the same, and the curve of SD locates in-between of TD and FD/CD, as given in~\eqref{eqnLatencyTD}--\eqref{eqnLatencySD}. Note that here we plot the average CDL of FD and CD in one curve as the values of beam duration $t_\text{FD}$ and $t_\text{CD}$ are exactly the same. Different behaviors are experience in Fig.~\ref{Figure8:b} where in terms of OH, SD outperforms all schemes as demonstrated in~\eqref{eqnOHSD}. It is worth nothing that in Fig.~\ref{Figure8:b}, the reason that TD/FD/CD performs the same in OH is the assumption that $\text{GI}=0$, where~\eqref{eqnOHFDCD} can be reformed as
\begin{align*}
\text{OH}_\text{FD} (= \text{OH}_\text{\text{CD}}) &= \frac{t_\text{FD/CD} S_{\text{FD/CD}}}{T} = \frac{\frac{N}{M} \cdot Mt_\text{TD} }{T} \\
&= \text{OH}_\text{\text{TD}}. \numberthis \label{eqnOHFDCDNew}
\end{align*}
In case $\text{GI}\neq0$, highest OH can be expected from TD as indicated in~\eqref{eqnOHFDCD}.

\subsubsection*{If so, how many simultaneous beams should be exploited (how to select $M$)?}~\\
From Fig.~\ref{Figure8:a}, the CDL increases with the number of simultaneous beams (the latency keeps fixed for TD as only $1$ beam is exploited). By contrast, the OH degrades with the increase of $M$ as indicated in Fig.~\ref{Figure8:b}. Therefore, the results in Fig.~\ref{Figure8} do not recommend any optimal simultaneous beam numbers $M$, unless the targeted performance metric is explicitly stated. Nevertheless, if both latency and OH are to be considered, SD provides the flexibility to achieve a trade-off between both metrics. In other words, by configuring the number of simultaneous beams, latency can be traded with OH, or vice versa.

\begin{figure}[htb]
	\centering
	\begin{subfigure}{0.48\textwidth}
		\includegraphics[scale=0.55]{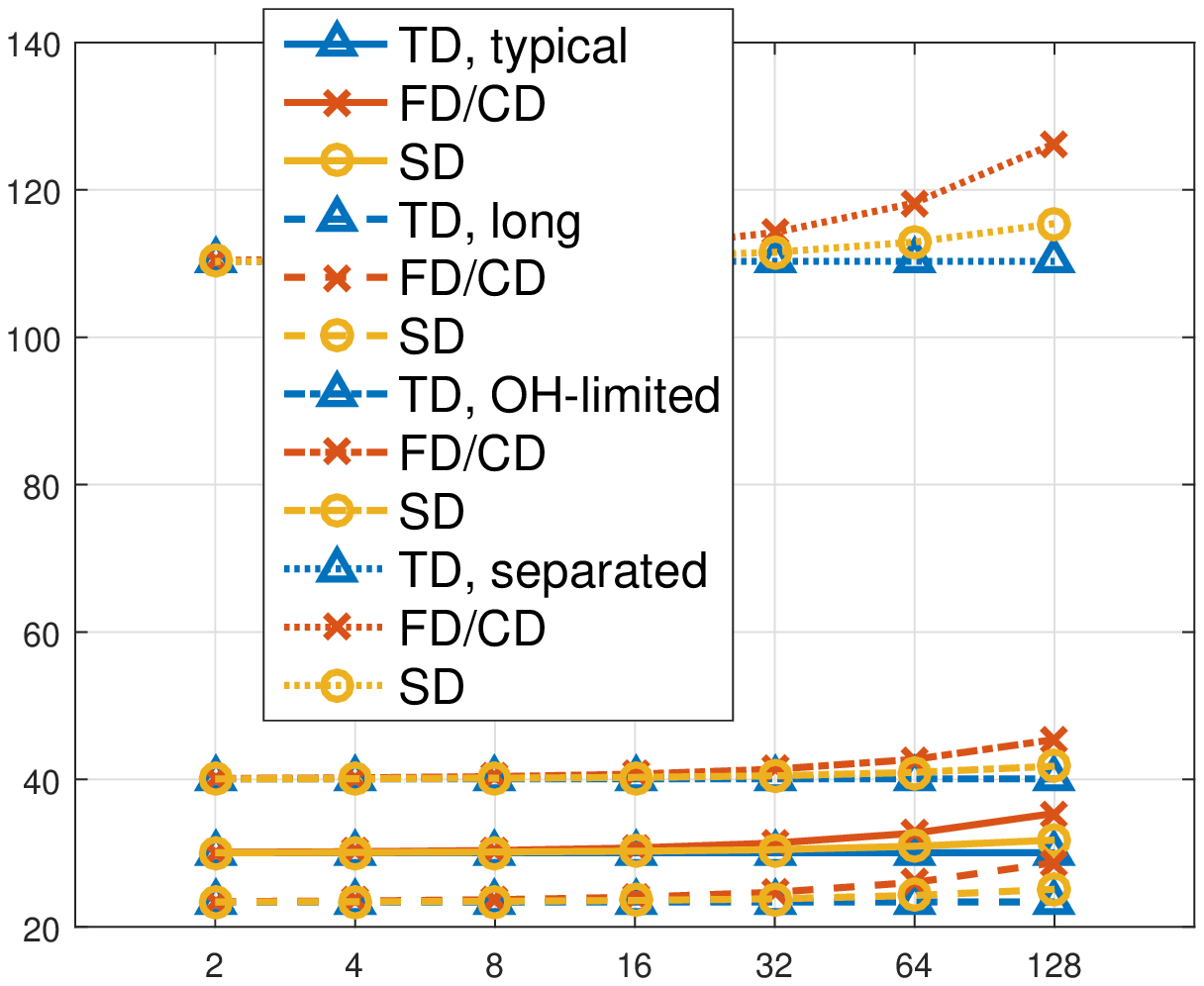}
		\caption{Latency}
		\label{Figure9:a}
	\end{subfigure}
	~ 
	\begin{subfigure}{0.48\textwidth}
		\includegraphics[scale=0.55]{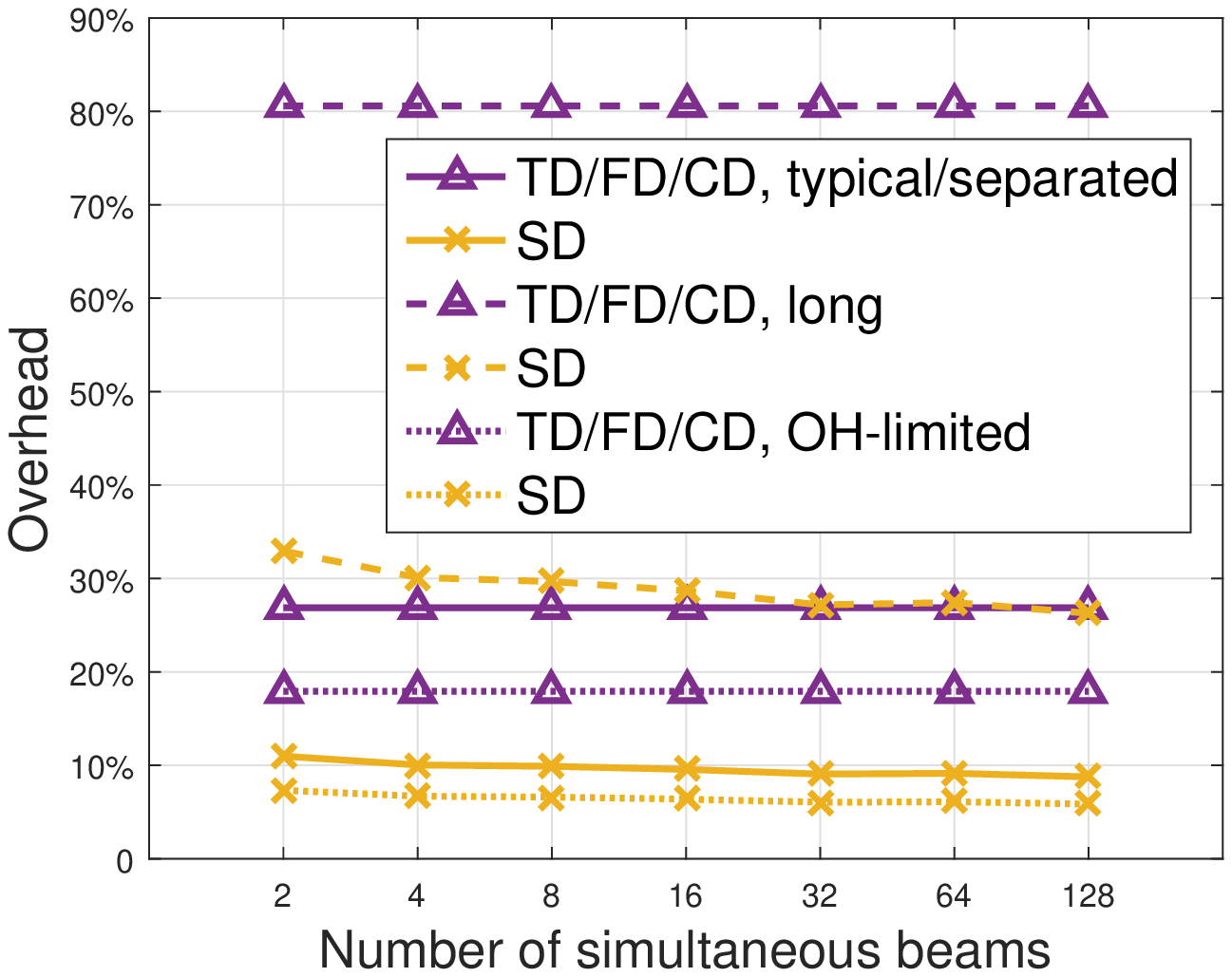}
		\caption{Overhead}
		\label{Figure9:b}
	\end{subfigure}
	\caption{Performance of different broadcast signaling schemes for cell discovery without knowledge of beacon timing and with completely ($W=3$), generally ($V=2$), and separately ($X=3$) integrated beacon interval are compared in (a) average CDL and (b) OH.}\label{Figure9}
\end{figure}

Fig.~\ref{Figure9} plots the average CDL and the OH of different broadcast signaling schemes without knowledge of beacon timing and with completely ($W=3$), generally ($V=2$), and separately ($X=3$) integrated beacon interval, defined in~\eqref{eqn20}--\eqref{eqn22}, versus the number of simultaneous beams $M$. The number of beam scan areas $N$ is also assumed as $128$.

The results in Fig.~\ref{Figure9} suggest that the average CDL declines when long frame is incorporated, while decrease in the latency leads to triple OH. On the other hand, for OH-limited and separated beacon interval cases, increased latencies are observed as costs for one-third less and identical OHs. Note that the less and identical OHs in the OH-limited case and in the separated beacon interval case, respectively, refer to superframe-scale. In the context of frame-scale, the OH of the typical frame for OH-limited case remains as the original, and the OH of data-only frame is $0$. For separated beacon interval case, the OH is $X$ time less as the original.

From the results in Fig.~\ref{Figure8} and Fig.~\ref{Figure9}, we conclude that the baseline broadcast signaling scheme TD, which is the single beam exhaustive scan, achieves the lowest CDL. By contrast, TD, as well as FD and CD, suffer from high OH. If both latency and OH are to be considered, multi-beam simultaneous scan (SD) provides the flexibility to achieve a trade-off between both performance metrics. In other words, by configuring the number of simultaneous beams, latency can be traded with OH, or vice versa. Furthermore, the trade-off can be extended to specific frame structure design, in which long frame with multiple beacon intervals results in lower latency at the price of higher OH. On the contrary, in the case of OH sensitive application scenario, superframe with embedded data-only frame, and/or equally beacon interval separation, are recommended to achieve lower OH with relative high latency.

\subsection{Cell Discovery Performance Comparison for Different Block Error Rates}
\subsubsection*{What is the impact of block error rate (how to select $\epsilon$)?}
\begin{figure}[htb]
	\centering
	\begin{subfigure}{0.48\textwidth}
		\includegraphics[scale=0.55]{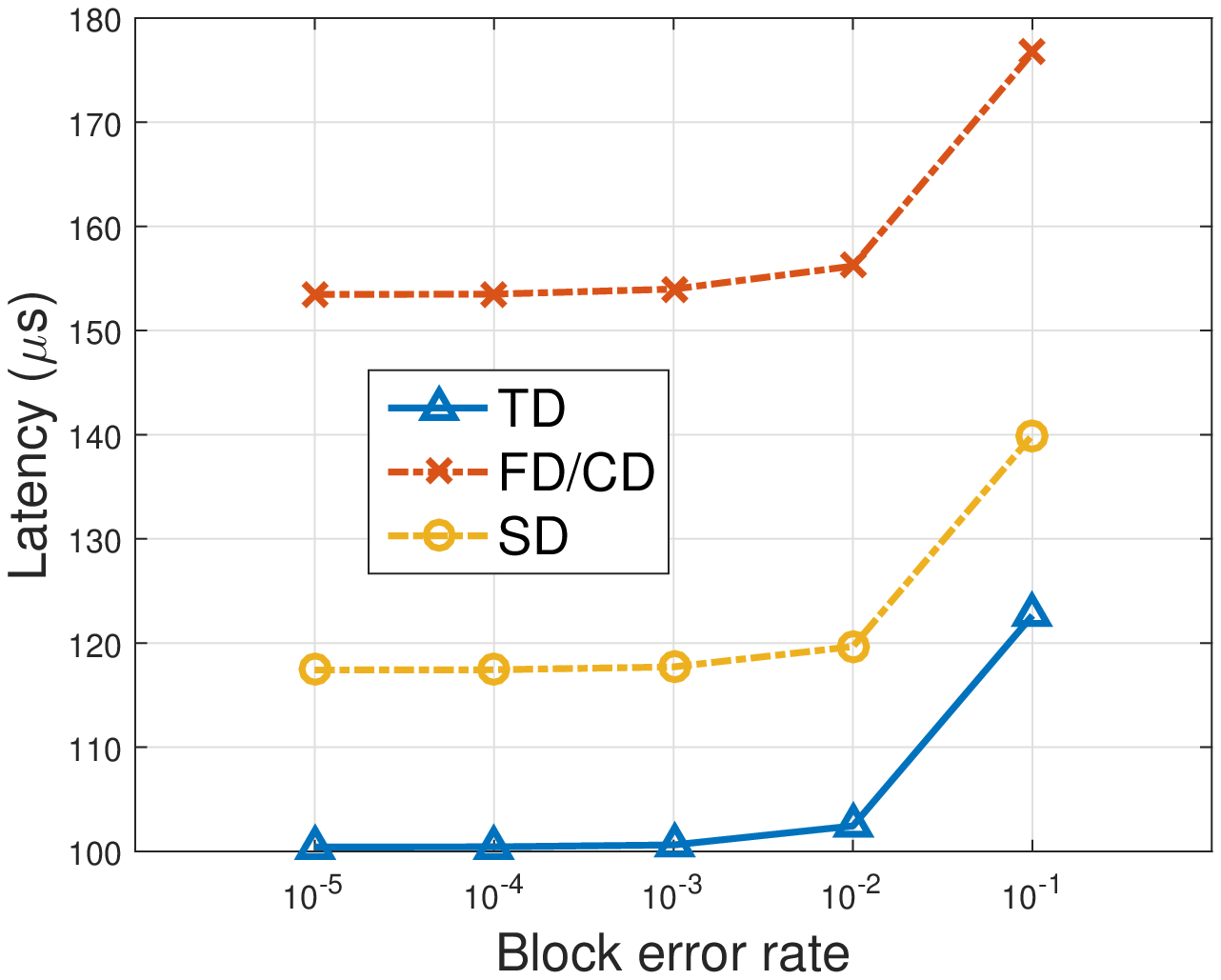}
		\caption{Latency}
		\label{Figure10:a}
	\end{subfigure}
	~ 
	\begin{subfigure}{0.48\textwidth}
		\includegraphics[scale=0.55]{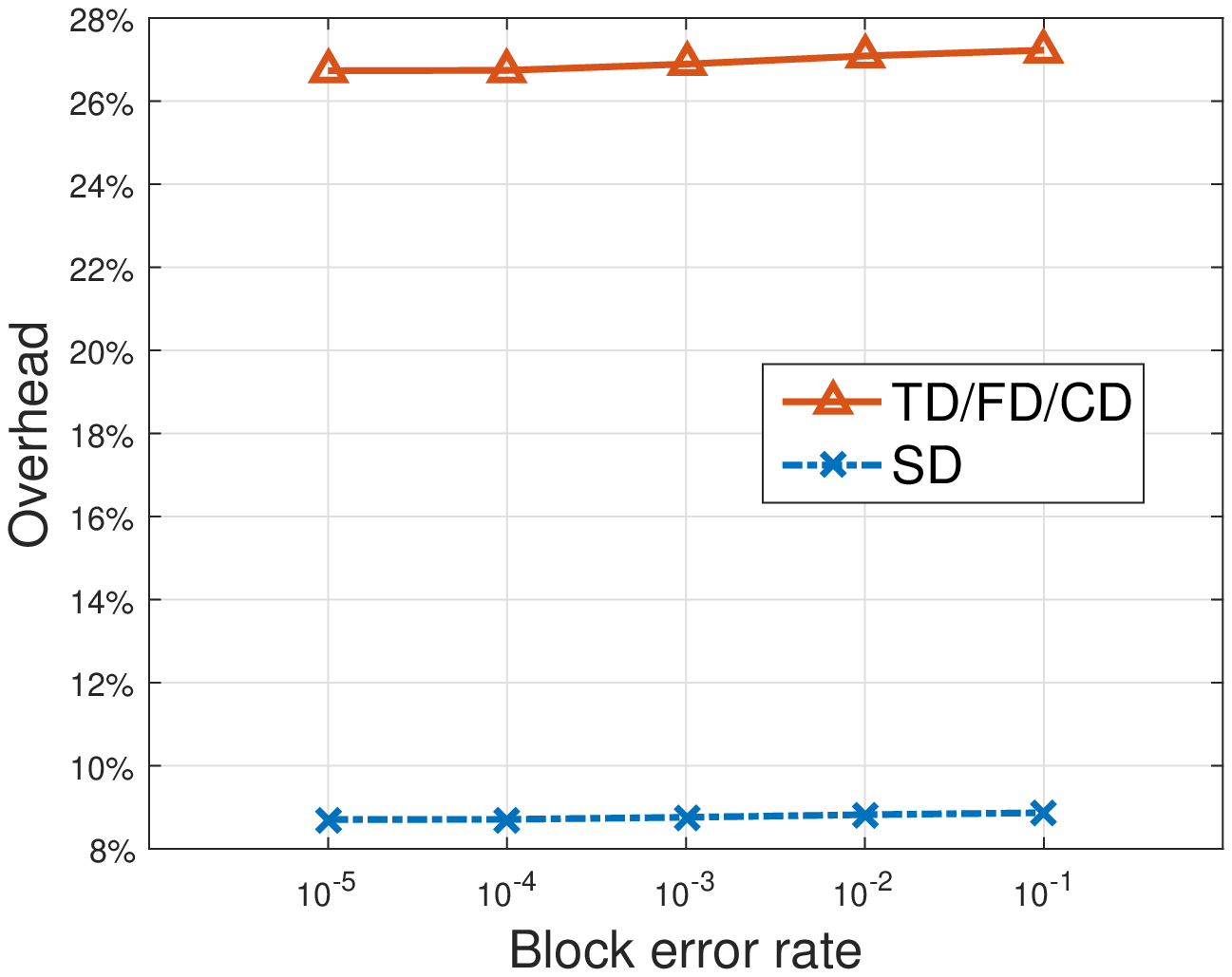}
		\caption{Overhead}
		\label{Figure10:b}
	\end{subfigure}
	\caption{Performance of different broadcast signaling schemes for different block error rates are compared in terms of (a) average CDL and (b) OH.}\label{Figure10}
\end{figure}

~\\
In Fig.~\ref{Figure10:a} and Fig.~\ref{Figure10:b}, the average CDL and the OH of different broadcast signaling schemes versus the block error rates $\epsilon$ are compared. 

In the figure, we notice that applying codes with low block error rate yields both lower CDL and OH. The main reason behind this behavior is the impact of low block error rate on the SNR, which decreases the beam duration $t$. However, these results show that the CDL and the OH are relatively insensitive to extreme low block error rates ($10^{-5}\text{--}10^{-3}$). Therefore, the relative high block error rate ($10^{-3}$) would be sufficient for initial cell discovery, unless the extreme coding scheme is desired to achieve the best performance.

\section{Findings and Conclusions}
In this paper, we proposed an analytical framework to investigate the performance of broadcast signaling design for mm-wave beamformed cell discovery. Specifically, we analyzed four schemes for the broadcast signaling where an AP delivers information to UEs for the cell discovery. Based on this, the cell discovery analyses were distinguished by beacon timing, scenarios, and beacon interval integration. By evaluating the performance metrics including cell discovery latency and signaling overhead, this paper allow us to make the following performance insights:

\begin{itemize}
	\item \textit{How wide should the beam be?} The cell discovery latency is optimized when the thinnest beam is formed. Interestingly, the beamforming architecture has no impact on the cell discovery latency, which makes the performance of analog/hybrid beamforming as well as the digital beamforming in terms of the cell discovery latency. By contrast, thiner beam results in higher signaling overhead.
	\item \textit{Is it beneficial to exploit multi-beam simultaneous scan?} Multi-beam simultaneous scan leads to cell discovery latency penalty. The baseline exhaustive scan scheme is found to be optimal in terms of the cell discovery latency. This is reversed when considering the signaling overhead, where the baseline exhaustive scan scheme, as well as the frequency-division/code-division scan scheme, suffer from high signaling overhead. The spatial-division scan scheme, however, achieves the lowest signaling overhead.
	\item \textit{If so, how many simultaneous beams should be exploited?} On the one hand, the cell discovery latency get worse as the number of simultaneous formed beams increases. On the other hand, the signaling overhead degrades with more simultaneous formed beams. The optimal number of simultaneous beams depends on the targeted performance metric. Nevertheless, the best trade-off between the cell discovery latency and the signaling can be achieved by the spatial-division beam scan. In other words, by configuring the number of simultaneous beams, latency can be traded with overhead, or vice versa.
	\item \textit{What is the impact of block error rate?} It has been demonstrated that the cell discovery latency and the signaling overhead are relatively insensitive to extreme low block error rates ($10^{-5}\text{--}10^{-3}$). Therefore, the relative high block error rate  ($10^{-3}$) would be sufficient for initial cell discovery, unless the extreme coding scheme is desired to achieve the best performance.
\end{itemize}

Future work can leverage the proposed analytical framework to investigate the performance of the cell discovery considering the beamformed single- or multi-beam scan on the UE side, or to extend the analytical framework to the complete initial access procedure. It would also be interesting to include various system models, e.\,g., to address other types of spatial locations, blockage models, and/or fading channels.

\appendices
\section{Discussion of Orthogonal Codes for Code-Division Broadcast Signaling Scheme}
From an information theoretical point of view, code-division is completely equivalent to time-frequency division when orthogonal codes, which do not lose orthogonality
after transmission due to multi-path channel, are utilized. In other words, for a frequency flat channel, achieving orthogonality in time, frequency, or in any other multiplexing dimension, is ``exactly'' the same in terms of spectral efficiency and SNR, therefore orthogonal code-division multiple access (CDMA) performs exactly the same as time-division multiple access (TDMA) or frequency-division multiple access (FDMA). 

A standard direct spreading CDMA system typically correlates the signal with respect to the desired spreading code, and treats the rest as additive noise. In this case, CDMA is a simplistic way to achieve the low rate coding~\cite{Viterbi}. In short, it consists of concatenating a given channel code with a repetition code of rate $\frac{1}{SG}$, where $\text{SG}$ is the spreading gain. 

Non-orthogonal codes may slightly benefit from special sets of sequences with particularly good correlation coefficients, however in this case, it makes more sense to use orthogonal codes. When considering the robustness to multi-path for non-orthogonality, it has been shown in Qualcomm IS-95 and Wideband CDMA (WCDMA) that random spreading is as good as any other family of sequences. In this case, the receive power after de-spreading is attenuated by a factor $\frac{1}{SG}$, and the approach for the Gaussian channel in~\eqref{eqnt} is still valid, taking into account the overhead of SG dimensions per symbol due to spreading. Moreover, utilization of directional antenna leads to relatively low multi-user interference where more simultaneous transmissions can be supported to exploit spatial multiplexing gain. When the interference is weak, imposing decoding or cancellation of all signals, instead of just the useful one, is not a good strategy.

In conclusion, insisting on treating interference as noise, it is beneficial to consider the orthogonal codes, which is strictly better than any form of random spreading, then we have the family of orthogonal access, namely TDMA, FDMA, and orthogonal CDMA, which are all equivalent and yield exactly the same SNR and channel capacity performance. 

\section{Derivation of Average Cell Discovery Latency without Knowledge of Beacon Timing and Single Complete Integrated Beacon}
Derivation of $\overline{T}_\mathrm{A}$ in~\eqref{eqn3}: 
\begin{align*}
\overline{T}_\mathrm{A} &= \sum_{i=1}^{S-1} p_\mathrm{A}(i) t_\mathrm{A}(i) \\
&= \sum_{i=1}^{S-1} \frac{t'}{T} \frac{S-i}{S} \Bigg( \bigg( \frac{t'}{2} + \frac{S-i-1}{2}t'  + t \bigg)(1-\epsilon) \\
& \quad + \sum_{k=1}^{K-1} \bigg( \frac{t'}{2} + T - it' + (k-1)T + \frac{S+i-1}{2}t' + t \bigg) \epsilon^k(1-\epsilon) + \bigg( \frac{t'}{2} + T - it' + (K-1)T \bigg) \epsilon^K \Bigg)\\
&= \sum_{i=1}^{S-1} \frac{t'}{T} \frac{S-i}{S} \Bigg( \bigg(\frac{S-i}{2}t' + t \bigg)(1-\epsilon) + \sum_{k=1}^{K-1} \bigg( kT + \frac{S-i}{2}t' + t \bigg) \epsilon^k(1-\epsilon) + \bigg(KT + \bigg(\frac{1}{2} - i\bigg)t' \bigg) \epsilon^K \Bigg) \\
&= \sum_{i=1}^{S-1} \frac{t'}{T} \frac{S-i}{S} \Bigg( \bigg(\frac{S-i}{2}t' + t \bigg)(1-\epsilon) + \frac{\epsilon+\epsilon^{K+1}(K-1)-K\epsilon^K}{1-\epsilon}T + (\epsilon-\epsilon^K)\bigg(\frac{S-i}{2}t' + t \bigg) \\
&\quad + \bigg(KT + \bigg(\frac{1}{2} - i\bigg)t' \bigg) \epsilon^K \Bigg). \\
\end{align*}
Combing the terms including $i$, we have
\begin{align*}
\overline{T}_\mathrm{A} &= \sum_{i=1}^{S-1} \bigg(\frac{t'}{T} - \frac{t'}{TS}i\bigg) \Bigg( (1-\epsilon^K) \bigg(t+\frac{S}{2}t'\bigg) + \frac{\epsilon^K}{2}t' + \frac{\epsilon-\epsilon^{K+1}}{1-\epsilon}T - \frac{1+\epsilon^K}{2}t'i  \Bigg).
\end{align*}
Let $A_1= (1-\epsilon^K) \big(t+\frac{S}{2}t'\big) + \frac{\epsilon^K}{2}t' + \frac{\epsilon-\epsilon^{K+1}}{1-\epsilon}T$, $A_2= \frac{1+\epsilon^K}{2}t' $, then
\begin{align*}
\overline{T}_\mathrm{A} &= \frac{t'A_1}{T}\sum_{i=1}^{S-1}1 - \bigg(\frac{t'A_1}{TS}+\frac{t'A_2}{T} \bigg) \sum_{i=1}^{S-1}i + \frac{t'A_2}{TS}\sum_{i=1}^{S-1}i^2 \\
&= \frac{(S-1)A_1}{T}t' - \frac{(S-1)A_1}{2T}t' - \frac{S(S-1)A_2}{2T}t' + \frac{(S-1)(2S-1)A_2}{12T}t'\\
&= \frac{(S-1)A_1}{2T}t' - \frac{(S-1)(4S+1)A_2}{12T}t'. \numberthis \label{eqn15}
\end{align*}

Derivation of $\overline{T}_\mathrm{B}$ in~\eqref{eqn6}:
\begin{align*}
\overline{T}_\mathrm{B} &= \sum_{i=1}^{S} p_\mathrm{B}(i) t_\mathrm{B}(i) \\
&= \sum_{i=1}^{S-1} \frac{t'}{TS} \bigg( \frac{t'}{2} + T - it' +  \sum_{k=0}^{K-2} \big( kT + t  + (i-1)t' \big) \epsilon^k(1-\epsilon) + (K-1)T\epsilon^{K-1} \bigg)\\
&= \sum_{i=1}^{S} \frac{t'}{TS} \bigg( \frac{t'}{2} + T - it' + (1-\epsilon^{K-1})\big(t  + (i-1)t'\big) + \frac{\epsilon+\epsilon^{K}(K-2)-\epsilon^{K-1}(K-1)}{1-\epsilon}T + (K-1)T\epsilon^{K-1} \bigg) \\
&= \sum_{i=1}^{S} \frac{t'}{TS} \bigg( (1-\epsilon^{K-1}) (t-t') + \frac{t'}{2} + \frac{1-\epsilon^{K}}{1-\epsilon}T - \epsilon^{K-1}t'i \bigg).
\end{align*}
Let $B_1= (1-\epsilon^{K-1}) (t-t') + \frac{t'}{2} + \frac{1-\epsilon^K}{1-\epsilon}T$, $B_2= \epsilon^{K-1}t' $, then
\begin{align*}
\overline{T}_\mathrm{B} &= \frac{t'B_1}{TS}\sum_{i=1}^{S}1 - \frac{t'B_2}{TS} \sum_{i=1}^{S}i \\
&= \frac{B_1}{T}t' - \frac{(S+1)B_2}{2T}t'. \numberthis \label{eqn16}
\end{align*}

The derivation of $\overline{T}_\mathrm{C}$ in~\eqref{eqn9} and $\overline{T}_\mathrm{D}$ in~\eqref{eqn11} can be done similarly as the in~\eqref{eqn16}. Due to space limitation, we refrain from going into more details of these derivations, and provide the results as follows:

Let $C_1= (1-\epsilon^{K-1}) (t-t') + \frac{t'}{2} + \frac{1-\epsilon^K}{1-\epsilon}T$, $C_2= \frac{1+\epsilon^{K-1}}{2}t' $, then
\begin{align*}
\overline{T}_\mathrm{C} &= \frac{(S-1)(S+2)(C_1+C_2)}{TS}t' - \frac{(S-1)C_1}{2TS}t' - \frac{S(S+1)(2S+1)C_2-C_2}{6TS}t'. \numberthis \label{eqn17}
\end{align*}

\begin{align*}
\overline{T}_\mathrm{D}
&= \frac{T-St'}{T} \Bigg( (1-\epsilon^{K-1})t - \frac{(S-1)\epsilon^{K-1}+1}{2}t' + \frac{\frac{1+\epsilon}{2}-\epsilon^K}{1-\epsilon}T \Bigg). \numberthis \label{eqn18}
\end{align*}

When $K\rightarrow \infty$, we have
\begin{align*}
A_1 &= t+\frac{S}{2}t' + \frac{\epsilon}{1-\epsilon}T, \\
A_2 &= \frac{t'}{2},\\
B_1 &=  t-t' + \frac{t'}{2} + \frac{T}{1-\epsilon} = t - \frac{t'}{2} + \frac{T}{1-\epsilon}, \\
B_2 &= 0,\\
C_1 &= B_1 = t - \frac{t'}{2} + \frac{T}{1-\epsilon}, \\
C_2 &= \frac{t'}{2}, \\
\overline{T}_\mathrm{D} &= \frac{T-St'}{T} \Bigg( t - \frac{t'}{2} + \frac{1+\epsilon}{2(1-\epsilon)}T \Bigg). \numberthis \label{eqn19}
\end{align*}
Taking~\eqref{eqn19} back to~\eqref{eqn15}-\eqref{eqn18}, we can get the results in~\eqref{eqn14}. Due to space limitation, we refrain from going into more details of the derivation of $\overline{T}_\text{w/o}$.

\bibliographystyle{IEEEtran}
\bibliography{mm_wave}

\end{document}